\documentclass[usenatbib]{mnras}
\usepackage{natbib,amssymb,amsmath,times,graphicx,url,aas_macros}
\usepackage{setspace}      
\usepackage{epstopdf}       
\usepackage{verbatim}       
\usepackage{bm}		  
\usepackage{fleqn}		  
\usepackage{Wurster_macros}  

\newcommand{\fid}{$fid$}
\newcommand{\fidm}{$fid_{3\text{e}5}$}
\newcommand{\fidlish}{$fid_{1\text{e}5}$}
\newcommand{\fidl}{$fid_{3\text{e}4}$}
\newcommand{\mhalf}{$M_{0.5}$}
\newcommand{\mtwo}{$M_2$}

\newcommand{\mangvlow}{$\theta_{15}$}
\newcommand{\manglow}{$\theta_{30}$}
\newcommand{\manghigh}{$\theta_{60}$}
\newcommand{\mfast}{$v_\text{fast}$}
\newcommand{\mslow}{$v_\text{slow}$}
\newcommand{\mmachvhigh}{$\beta_\text{8}$}
\newcommand{\mmachhigh}{$\beta_\text{2}$}
\newcommand{\mmachlow}{$\beta_\text{0.5}$}

\newcommand{\vx}{$v_\text{x}$}
\newcommand{\vy}{$v_\text{y}$}

\newcommand{\vxi}{$v_\text{x,0}$}
\newcommand{\vyi}{$v_\text{y,0}$}

\title[Gas and star kinematics in cloud-cloud collisions]{Gas and star kinematics in cloud-cloud collisions}

\author[Wurster \& Bonnell]{James Wurster$^{1}$\thanks{jhw5@st-andrews.ac.uk} and Ian A. Bonnell$^{1}$\\
$^{1}$Scottish Universities Physics Alliance (SUPA), School of Physics and Astronomy, University of St. Andrews, North Haugh, St Andrews, Fife KY16 9SS, UK \\
}
\date{Submitted: Revised: Accepted: }
\pagerange{\pageref{firstpage}--\pageref{lastpage}} \pubyear{2021}

\begin{document}
\label{firstpage}
\bibliographystyle{mnras}
\maketitle

\begin{abstract}
We model the collision of molecular clouds to investigate the role of the initial properties on the remnants.  Our clouds collide and evolve in a background medium that is approximately ten times less dense than the clouds, and we show that this relatively dense background is dynamically important for the evolution of the collision remnants.  Given the motion of the clouds and the remnants through the background, we develop, implement, and introduce dynamic boundary conditions.  We investigate the effect of the initial cloud mass, velocity, internal turbulence, and impact angle.  The initial velocity and its velocity components have the largest affect on the remnant.  This affects the spatial extent of the remnant, which affects the number of resulting star clusters and the distribution of their masses.  The less extended remnants tend to have fewer, but more massive, clusters.  Unlike the clusters, the gas distributions are relatively insensitive to the initial conditions, both the distribution of the bulk gas properties and the gas clumps.  In general, cloud collisions are relatively insensitive to their initial conditions when modelled hydrodynamically in a dynamically important background medium. 
\end{abstract}

\begin{keywords}
stars: formation --- ISM: clouds ---galaxies: star clusters: general --- methods: numerical.
\end{keywords} 

\section{Introduction}
\label{intro}

The Galaxy is a dynamic environment, populated by molecular clouds that collide every \sm10~Myr or so \citepeg{TaskerTan2009,Tasker2011,DobbsPringleDuartecabral2015}.  
Many collisions have been observed, including 
Westerlund 2 \citep{Furukawa+2009,Ohama+2010}, 
NGC 3603 \citep{Fukui+2014}, 
G0.253+0.016 \citep{Higuchi+2014}, 
RCW 38 \citep{Fukui+2016}, 
R136 \citep{Fukui+2017}, 
GM 24 \citep{Fukui+2018gm}, 
M42 \& M43 \citep{Fukui+2018orion}, 
M 33 \citep{Sano+2021}, 
NGC 2023 \citep{Yamada+2021}, and 
G31.41+0.31 \citep{Beltran+2022}.
For a review, see \citet{Fukui+2021}.  
These observations suggest that the collisions are typically head-on between clouds of different sizes and masses at speeds of a few 10s of \kms{}.  Although less commonly observed, there are observations of clouds that collide at oblique angles \citep[e.g., NGC 2068 and NGC 2071;][]{Fujita+2021}.

One identifying feature of a (head-on) cloud-cloud collision is a `bridge' that appears in position-velocity space \citep{Haworth+2015bridgeA,Haworth+2015bridgeB}; however, these features are transitory and their observability depends on viewing angle.  Nonetheless, this bridge has been identified both observationally and numerically.  Bridging features are typically explored in CO, however, at very low densities, it is possible that there would not be enough CO emission to trace the bulk flows of the clouds \citep{Clark+2019}. At higher densities, molecules that trace the dense gas, such as NH$_3$ and HCN, have enhanced emission that serve to highlight the CO bridging features \citep{PriestleyWhitworth2021}.

One motivation for investigating cloud-cloud collisions is that they are expected to be the site of high-mass star formation; this is reinforced by observations such as those listed above.   Indeed, these observations further suggest that colliding clouds can trigger (high-mass) star formation, particularly when the clouds collide head-on with fast impact velocities  \citepeg{Fukui+2021}.  Another possible path to high-mass star formation is first through the formation of hub-filament networks  \citepeg{Myers2009,Balfour+2015,BalfourWhitworthHubber2017,Beltran+2022}, where the filaments are generally expected to feed the hubs.  It is then expected that these hubs are the location of high-mass star formation \citepeg{Kumar+2020}, which has been supported by observations of high-mass clumps in hubs \citepeg{Peretto+2013,Anderson+2021}.   Although hub-filament networks appear in simulations of cloud-cloud collisions, it should be noted that they also appear in numerical simulations that include driven \citepeg{FederrathKlessen2012,FederrathKlessen2013,TriccoPriceFederrath2016} or decaying \citepeg{Bate2012,WursterBatePrice2019} turbulence that do not explicitly model colliding clouds.  However, these scenarios can be mutually consistent since colliding flows generate turbulence \citepeg{Vazquezsemadeni+2007}.   Ultimately, through studying cloud-cloud collisions, we will hopefully be able to derive the elusive theory of high-mass star formation \citerev{ZinneckerYorke2007} and resolve the competing theories of monolithic collapse \citep{MckeeTan2003} versus competitive accretion \citep{Bonnell+1997,Bonnell+2001ca} versus inertial inflow \citep{Padoan+2020}. 

Given their importance for high-mass star formation, there have been numerous numerical studies of colliding flows and cloud-cloud collisions using a plethora of initial conditions.  Several studies employ the prescribed cooling curve of \citet{KoyamaInutsuka2002}, where the gas is typically initialised in an unstable thermal equilibrium at $n \approx 1$~\percc{}.  Perturbations permit the gas to cool and collapse or to heat and disperse, creating a two-phase medium where the gas preferentially lies on the cooling curve.  At the shock interface between colliding flows/clouds, however, the gas is shock-heated off of the equilibrium curve, but subsequently cools until it equilibrates on the cooling curve \citepeg{KoyamaInutsuka2002,Vazquezsemadeni+2007,CarrollnellenbackFrankHeitsch2014,Fogerty+2016}.  During this process, cool dense clumps are formed, typically at/near the shocked regions.  Once the clumps have formed in the two-phase media, the surrounding warm gas cools and is accreted onto the clumps, indicating the importance of both media \citep{KoyamaInutsuka2002}.  Moreover, these clouds are not in virial equilibrium and undergo secular evolution before they collapse into stars  \citep{Vazquezsemadeni+2007}.  

When comparing the prescribed cooling curve of \citet{KoyamaInutsuka2002} to full non-equilibrium chemistry, \citet{Micic+2013} found that several cloud properties were insensitive to the cooling prescription (e.g., mass and volume filling factor), while others were dependent on the cooling prescription (e.g., cloud morphology and large-scale velocity distribution of the gas).  The gas distribution in the density-temperature phase space was crudely similar for each prescription, however, the temperature distribution was much broader with the non-equilibrium chemistry.  Therefore, the \citet{KoyamaInutsuka2002} cooling curve is a good and efficient approximation, but care must be applied when interpreting the results and comparing them to observations or other simulations where more complex physics is occurring. 

The interstellar medium is turbulent \citepeg{HeyerBrunt2004}, thus it is reasonable to expect that the colliding clouds/flows themselves are structured and turbulent prior to the collision.  Thus, for consistency and realism, numerical simulations tend to seed the clouds/flows with a (decaying) turbulent velocity field; this initial turbulence also helps to break the initial symmetry of the clouds/flows.  Unsurprisingly, the initial turbulence affects the results.  \citet{CarrollnellenbackFrankHeitsch2014} found morphological differences of their remnants when comparing initially smooth models to initially clumpy models; they also found that initially clumpy flows yielded more remnant clumps and more massive clumps, however, the onset of clump formation was delayed compared to the smooth initial flow.  There was also evidence for global collapse in the clumpy model that did not appear in the smooth model.  Turbulence is also generated through the collision itself \citepeg{Vazquezsemadeni+2007}, where the turbulence becomes compressive after the collision; however, if the initial velocity is too high, then the turbulence remains a compressive/solenoidal mix since the interaction occurs too quickly \citep{TakahiraTaskerHabe2014}.  

Star formation occurs in a magnetised medium \citerev{Crutcher2012}, therefore, many studies have modelled colliding clouds using magnetohydrodynamics \citepeg{Vazquezsemadeni+2011,InoueFukui2013,Fogerty+2016,DobbsWurster2021,Fukui+2021,Sakre+2021,KinoshitaNakamura2022}.  These simulations demonstrate the importance and impact of magnetic fields on the colliding clouds, with the results being dependent on both the magnetic field strength and orientation.  Nonetheless, \citet{DobbsWurster2021} found that magnetic fields do not impede the formation of clumps nor the development of high star formation rates.  Since our current study does not include magnetic fields, we refer the reader to the listed references for a discussion on magnetic fields and cloud-cloud collisions.

Numerical studies tend to model clouds colliding head-on or slightly off-set.  As described above, there is observational motivation for this setup, however, it is also numerically preferred since the computational domain can be well-defined prior to the simulation.  Collisions where there is a net bulk motion have yet to be numerically explored in part since the computational domain cannot be easily defined \textit{a priori}.  

In this paper, we model the hydrodynamic collision of two clouds with impact angles between 30$^\circ$ and 120$^\circ$ where the initial clouds and their remnant have a bulk forward motion.  The domain is initialised as a two-phase medium, therefore open boundaries are not possible.  Thus, for computational efficiency, in \appref{app:bdy}, we introduce dynamic boundaries that adapt as the clouds and remnant propagate; this prevents the needs to choose a domain \textit{a priori}.  The remainder of this paper is organised as follows.  In \secsref{sec:methods}{sec:ic}, we present our methods and initial conditions, respectively.  We present and discuss our results in \secref{sec:results}, and we conclude in \secref{sec:conc}.

\section{Methods}
\label{sec:methods}

Using the smoothed particle hydrodynamics (SPH) code \textsc{Phantom} \citep{Price+2018phantom}, we solve equations of self-gravitating hydrodynamics that are given by
\begin{flalign}
\frac{\text{d}\rho}{\text{d}t} & =  -\rho \nabla\cdot \bm{v}, \label{eq:cty} \\
\frac{\text{d} \bm{v}}{\text{d} t} & =  -\frac{1}{\rho}\bm{\nabla} P - \nabla\Phi,& \label{eq:mom} \\
\frac{\text{d} u}{\text{d}t} &=  -\frac{P}{\rho} \bm{\nabla} \cdot \bm{v} + \left.\frac{\text{d} {u}}{\text{d} t}\right|_\text{cool},&\label{eq:eng}\\
\nabla^{2}\Phi & =  4\pi G\rho,& \label{eq:grav} \\
P &= \left(\gamma -1\right)\rho u, &\label{eq:eos}
\end{flalign}
where $\frac{\text{d}}{\text{d}t} \equiv \frac{\partial}{\partial t} + \bm{v}\cdot\bm{\nabla}$ is the Lagrangian derivative,  $\rho$ is the density, ${\bm  v}$ is the velocity, $P$ is the gas pressure, $u$ is the internal energy, $\Phi$ is the gravitational potential, $G$ is the gravitational constant, and $\gamma (=5/3)$ is the adiabatic index.  
The cooling term is given by 
\begin{flalign}
\label{eq:dudtcool}
\left.\frac{\text{d} u}{\text{d} t}\right|_\text{cool} = \frac{1}{m_\text{H}} \Gamma - \frac{\rho}{m_\text{H}^2} \Lambda(T),
\end{flalign}
where $T$ is the gas temperature, $m_\text{H}$ is the mass of a hydrogen atom, and the heating and cooling terms are
\begin{subequations}
\label{eq:cool}
\begin{flalign}
\Gamma &= 2\times10^{-26} \text{erg \ s}^{-1}, &\\
\frac{\Lambda(T)}{\Gamma} &= \left[10^7 \exp\left(\frac{-118400}{T+1000}\right) + 0.014 \sqrt{T} \exp\left(\frac{-92}{T}\right) \right] \text{cm},
\end{flalign}
\end{subequations}
respectively \citep{KoyamaInutsuka2002,Vazquezsemadeni+2007}\footnote{\citet{KoyamaInutsuka2002} includes a typographical error which is corrected in \citet{Vazquezsemadeni+2007}; the correct form is used here.}.  The cooling terms are solved explicitly in \textsc{Phantom}, which requires a cooling timestep constraint \citep{GloverMaclow2007} of
\begin{flalign}
\label{eq:dtcool}
\text{d}t_\text{cool} = 0.3\left|\frac{u}{{\text{d} u}/{\text{d} t}}\right|.
\end{flalign}
To prevent $u < 0$, we impose a temperature floor of 3~K; see \appref{app:Tfloor}.  

Finally, to represent self-gravitating and collapsing high-density regions, we employ sink particles \citep{BateBonnellPrice1995}.  Sink particles are implemented when a candidate gas particle's density surpasses a critical density $\rho_\text{crit}$, and all the particles within twice its smoothing length pass the criteria described in \citet{BateBonnellPrice1995}.  We do not employ sink merging.  Bound regions that have yet to collapse into sinks are defined as clumps which are identified by our clump-finding algorithm introduced in \appref{app:clump}.

\section{Initial conditions}
\label{sec:ic}

The cooling curve given in Eqn.~\ref{eq:cool} defines the equilibrium temperature (or pressure) at any given density \citep[see also fig.~2 of][]{Vazquezsemadeni+2007}.  Therefore, we must choose initial densities and temperatures that are on this equilibrium curve\footnote{If we do not choose values on this curve, the gas equilibrates to this curve during the first few steps which requires short numerical timesteps.}.  Typical simulations that use this cooling curve \citepeg{KoyamaInutsuka2002,Vazquezsemadeni+2007} initialise a uniform medium with the unstable equilibrium density of $n \approx 1$~\percc{}.  In our study, however, we model the collision of two elliptical clouds travelling through a warm medium, thus we require two different densities on this curve that have the equal pressures.  We thus choose a density of $n = 3$~\percc{} for each cloud and $n = 0.33$~\percc{} for the background medium.  The cloud is given both a turbulent velocity (the implementation is as described in \citealp{OstrikerStoneGammie2001}, \citealp{BateBonnellBromm2003}, and \citealp{WursterBatePrice2019}) and a bulk velocity, where the bulk velocity, $v_0$ is parallel to the cloud's semi-major axis.  The clouds are rotated by an angle $\theta_0$ from the $x$-axis and separated such that the two clouds will impact at an angle of $2\theta_0$ after 2.2~Myr.   The background has a temperature of 6650~K, so that the background is in pressure equilibrium with the clouds.  To minimise the shock as the cloud moves into and through the background, the background velocity is $v_\text{bkg,x} = v_0 \cos\theta_0$ and $v_\text{bkg,y} = v_\text{bkg,z} = 0$.  While this minimises the shock in the $x$-direction, we cannot minimise the shock in the $y$-direction since the clouds are moving towards one another.

\begin{figure} 
\centering
\includegraphics[width=\columnwidth]{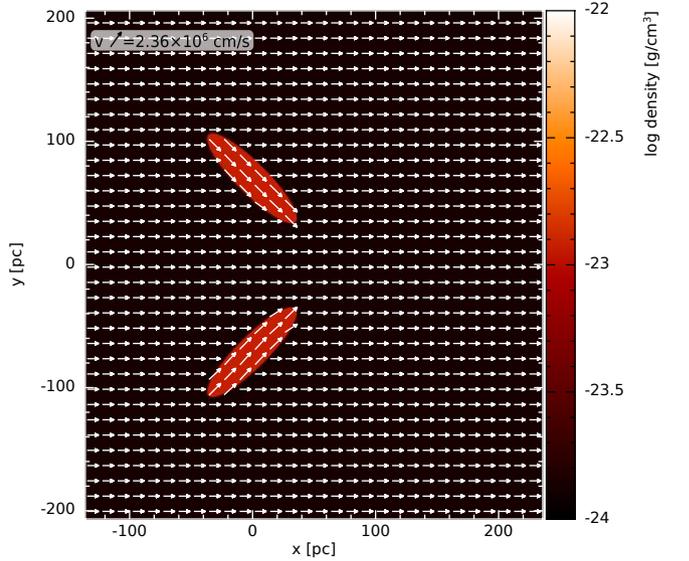}
\caption{Gas density in a cross section of the clouds in the initial state.  The arrows represent the initial velocity.  The cloud's velocity is a superposition of the bulk velocity parallel to its semi-major axis and the turbulent velocity.  The background gas has a bulk velocity $v_\text{bkg,x} = v_0 \cos\theta_0$ and $v_\text{bkg,y} = v_\text{bkg,z} = 0$ to minimise the shock between the interface between the cloud and background.} 
\label{fig:ics}
\end{figure}

The default properties of the cloud are given in \tabref{table:dflt} and the initial configuration is shown in \figref{fig:ics}.  Our default cloud size and ellipticity was selected to match the clouds identified in \citet{SmilgysBonnell2016}; as further motivation, elliptical clouds have also been observed \citepeg{Colombo+2014,ZuckerBattersbyGoodman2018}, observed in simulations \citepeg{DuartecabralDobbs2016}, and used in previous colliding cloud simulations \citepeg{LiowDobbs2020}.  However, clouds of this mass may be better representative of colliding clouds in interacting galaxies \citepeg{Matsui+2012,MatsuiTanikawaSaitoh2019}.  

Given our initial conditions, we do not know \textit{a priori} the size of the domain into which the remnants of the cloud collisions will expand.  To circumvent this, we develop and use dynamic boundaries which expand and contract as the simulation evolves.  This prevents needing to know the full domain in advance and prevents modelling numerous ambient background particles that are not (yet) important for the evolution of cloud collision.  See \appref{app:bdy} for details of the dynamic boundaries.  The boundaries themselves are periodic.  

We use $10^6$ equal-mass SPH particles in each cloud, and a dynamically changing number of equal-mass particles in the background medium.  Particles are initialised on a cubic lattice.

Sink particles of radius $r_\text{sink} = 0.25$~pc are inserted using a critical density of $\rho_\text{crit} = 10^{-20}$~\gpercc{}; the minimum resolved sink mass is \sm0.33~\Msun{}.  Even at this reasonably large radius, hundreds of sinks form in our simulations, each typically representing an entire star cluster.

\begin{table}
\centering
\begin{tabular}{l l}
\hline
parameter & value \\
\hline
semi-major axis & 50~pc \\
semi-minor axis & 12.5~pc \\
mass & 5780~\Msun \\
mass density & $1.2\times10^{-23}$~\gpercc \\
number density & 3~\percc \\
temperature & 730~K \\
$E_\text{turb}/E_\text{grav}$ & 1 \\
$v_0$ & 21.75~\kms{} \\
$\theta_0$ & 45$^\circ$ \\
\hline
\end{tabular}
\caption{The cloud parameters used for our fiducial model.  We assume a mean molecular mass of 2.381 when converting between mass density and number density. }
\label{table:dflt}
\end{table}

\subsection{Parameter space}
\label{sec:ic:ps}

Our suite includes 11 models, which are listed in  \tabref{table:suite}.  One primary property is changed per model, however, additional properties are changed for consistency.  For example, secondary properties are adjusted such that all clouds have the same initial density, the same mass resolution, and take 2.2~Myr to collide.  For the clouds with differing initial masses, the major and minor axes are modified by an equal ratio.
\begin{table}
\centering
\begin{tabular}{l l}
\hline
model name & primary property changed \\
\hline
\fid{} & --- \\
\mhalf{}           & $M = 0.5M_0 = 2890$~\Msun \\                       
\mtwo{}           & $M = 2M_0 = 11560$~\Msun \\                         
\mslow{}          & $v = \frac{2}{3}v_0 = 14.5$~\kms\\                 
\mfast{}            & $v = \frac{3}{2}v_0 = 32.6$~\kms\\                 
\mmachlow{}   & $E_\text{turb}/E_\text{grav} = 0.5$ \\               
\mmachhigh{}  & $E_\text{turb}/E_\text{grav} = 2$ \\                  
\mmachvhigh{} & $E_\text{turb}/E_\text{grav} = 8$ \\                  
\mangvlow       & $\theta = \frac{1}{3}\theta_0 = 15^\circ$ \\        
\manglow{}      & $\theta = \frac{2}{3}\theta_0 = 30^\circ$ \\         
\manghigh{}    & $\theta = \frac{4}{3}\theta_0 = 60^\circ$ \\          
\hline
\end{tabular}
\caption{Our suite of simulations.  The first column lists the model's name and the second column lists the primary property changed. }
\label{table:suite}
\end{table}

\begin{figure*}
\centering
\includegraphics[width=\textwidth]{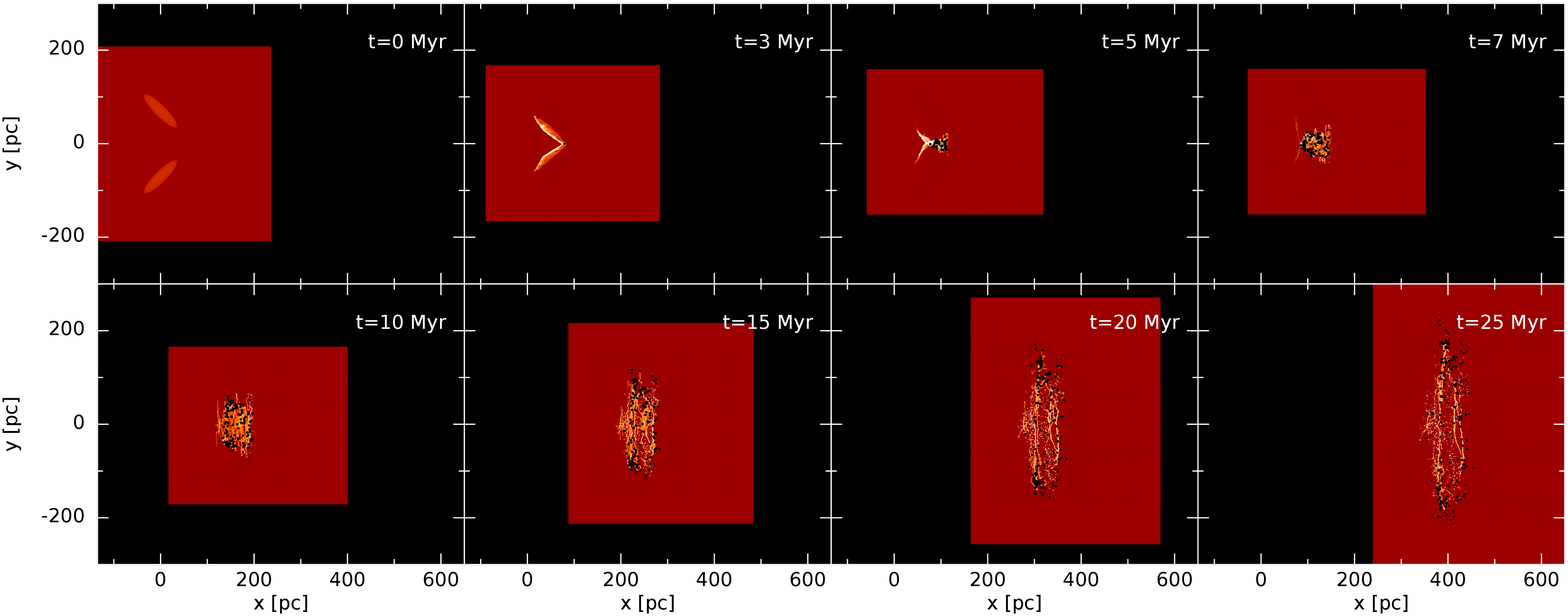}
\includegraphics[width=\textwidth]{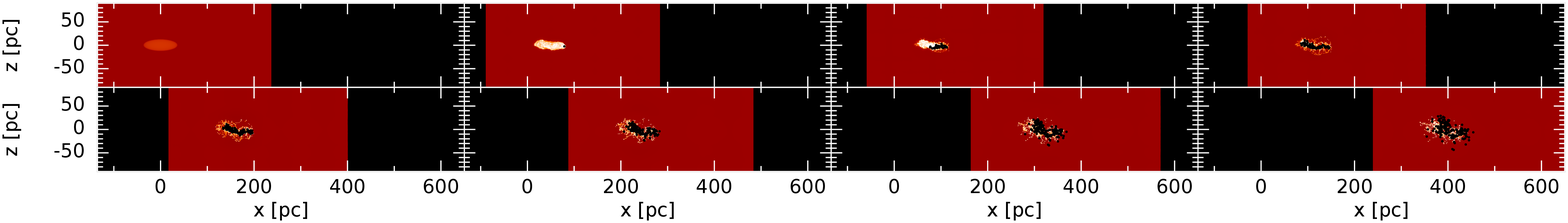}
\includegraphics[width=\textwidth]{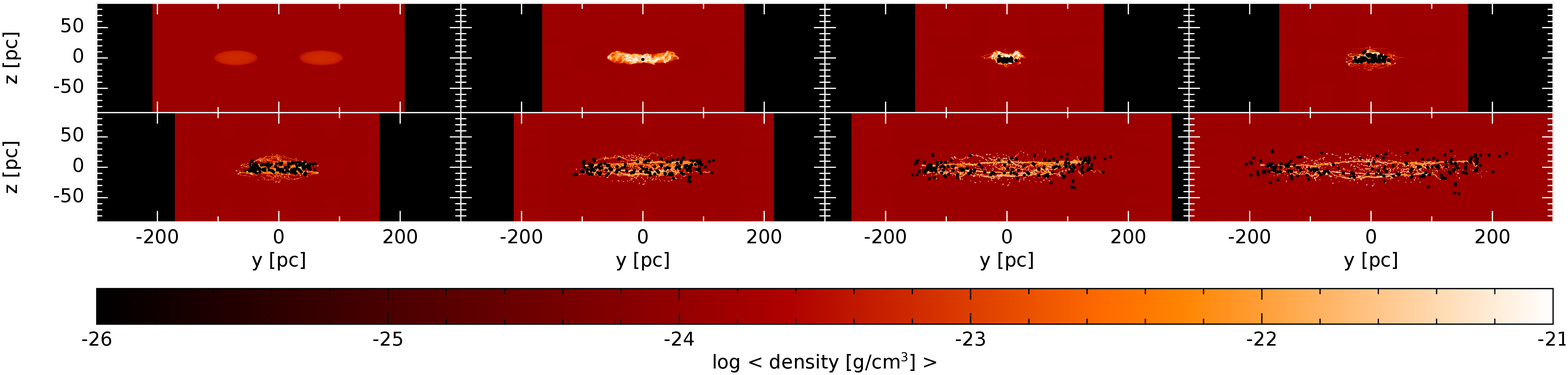}
\caption{Evolution of the average gas density of our fiducial model shown in three planes.  Times are listed in the top panel only for clarity.  Average gas density is used rather than column density since the column depth dynamically changes with the position of the boundaries in the third dimension.  Black represents the region outside of our computational domain as determined by the dynamical boundary conditions.  Black dots represent sink particles, plotted with a radius of 10x the actual sink radius for visualisation purposes; see \figref{fig:fid:evol:ns} for a version without sinks.  Clumps and filaments form during the merger, where the filaments continue to stretch out as they evolve.  Although the primary burst of star formation occurs during the collision, star formation continues in the filaments at a slower rate as they propagate.}
\label{fig:fid:evol}
\end{figure*} 

\section{Results}
\label{sec:results}
\subsection{General evolution}
\label{sec:ge}
The background medium is only a factor of \sm10 less dense than the initial clouds, thus it is dynamically important.  The medium provides additional material and applies additional forces on both the initial clouds and on the remnants of the collision as they move through the medium.  The evolution of our fiducial model is shown in \figref{fig:fid:evol}; a version excluding sink particles is shown in \figref{fig:fid:evol:ns} in \appref{app:figs}.

During the initial motion ($t \lesssim 3$~Myr), the clouds become asymmetric, with gas piling up on the trailing edge of the cloud and the leading edge becoming less dense.  This is despite the clouds and the background initially moving with the same $x$-velocity.  This pile up is a combination of shear between the cloud and the medium due to the differing initial $y$-velocities, the self-gravity of the cloud promoting gravitational collapse, and the initial turbulence of the cloud supporting against collapse.  The resulting asymmetric cloud suggests that the shear with the medium is the dominant process causing the asymmetry, thus the background performs its first dynamical role.  

The collision occurs between $t \sim 3 - 7$~Myr.  During this time, the two clouds merge into a small, but transient clump, accompanied by a burst of star formation (see \secref{sec:sd} below).  Moreover, this collision totally disrupts both initial clouds, well before there was any significant gravitational collapse.  

As the remnant evolves after the impact, it expands in all directions, but most notably in the $y$-direction.  The $y$-expansion is simply a result of the remnant retaining a memory of the initial velocity of the clouds, which was chosen to diverge after impact.  
However, based upon the initial velocity, the cloud was expected to reach $y_\text{final} \approx \pm 400$~pc, thus the collision and subsequent interactions with the background continually decrease \vy{} of the remnants to prevent this (see \secref{sec:gd:v} below).  Although the bulk remnant is slowing in this direction, it is not doing so uniformly, promoting the formation of filamentary structure in the $y$-direction.  

The expansion in the $x$- and $z$-directions is noticeably smaller than in the $y$-direction and is a result of the internal turbulence, both seeded and caused by the collision.  The total distance traversed in the $x$-direction is d$x \approx 400$~pc (\figref{fig:fid:evol}), which is the distance expected given the initial \vx{}.  Therefore, after the collision, the medium plays a reduced role in this direction, with all gas converging to a similar $v_\text{x}$ (see \secref{sec:gd:v} below); note that if $v_\text{bkg,x} = 0$, we would have expected a similar structure in the $x$-direction as seen in the $y$-direction.

After the collision, the clouds fragment into clumps (represented by both gas clumps and sinks) and filaments.  The filaments form and grow from large gas reservoirs that do not collapse to form stars, where the reservoir's velocity divergence (particularly the $y$-component) is enough to prevent the collapse and to promote the formation of the filament.  As the remnant evolves, the filaments become longer in the $y$-direction but thinner in the other two directions due to the small internal velocity dispersion in these directions.  This small velocity dispersion also accounts for the lack of filaments (or even low-density striations) in the $x$- or $z$-directions.   The filaments are moving with a similar \vx{} as the background medium, thus they are not growing substantially in mass by sweeping up much pristine gas (see \secref{sec:gd} below).  Although there is the clear formation of filaments, there is no observed hub-filament network \citep{Myers2009,Beltran+2022} of the classical radial hub-and-spoke type structure.  Moreover, these filaments are formed and maintained by the gas flow in the $\pm y$-direction, thus the gas is flowing away from the mid-plane rather that converging to any point (i.e., to a traditional hub).

When a gas clump becomes dense enough, it is replaced with a sink particle, where its initial velocity is the same as its progenitor clump; these progenitor clumps are typically over-densities in the filaments that are caused by local gravitational collapse rather than large-scale inflow along the filaments towards hubs.  Given this initial velocity and that sink particles do not feel gas pressure, they tend to detach from their birth region, and many move in advance of the expanding filaments and clumps into regions containing pristine background medium.  Therefore, these sinks grow by accreting both dense gas from other clumps and filaments and the pristine background gas (see \secref{sec:sd} below).

\begin{figure*} 
\centering
\includegraphics[width=0.75\textwidth]{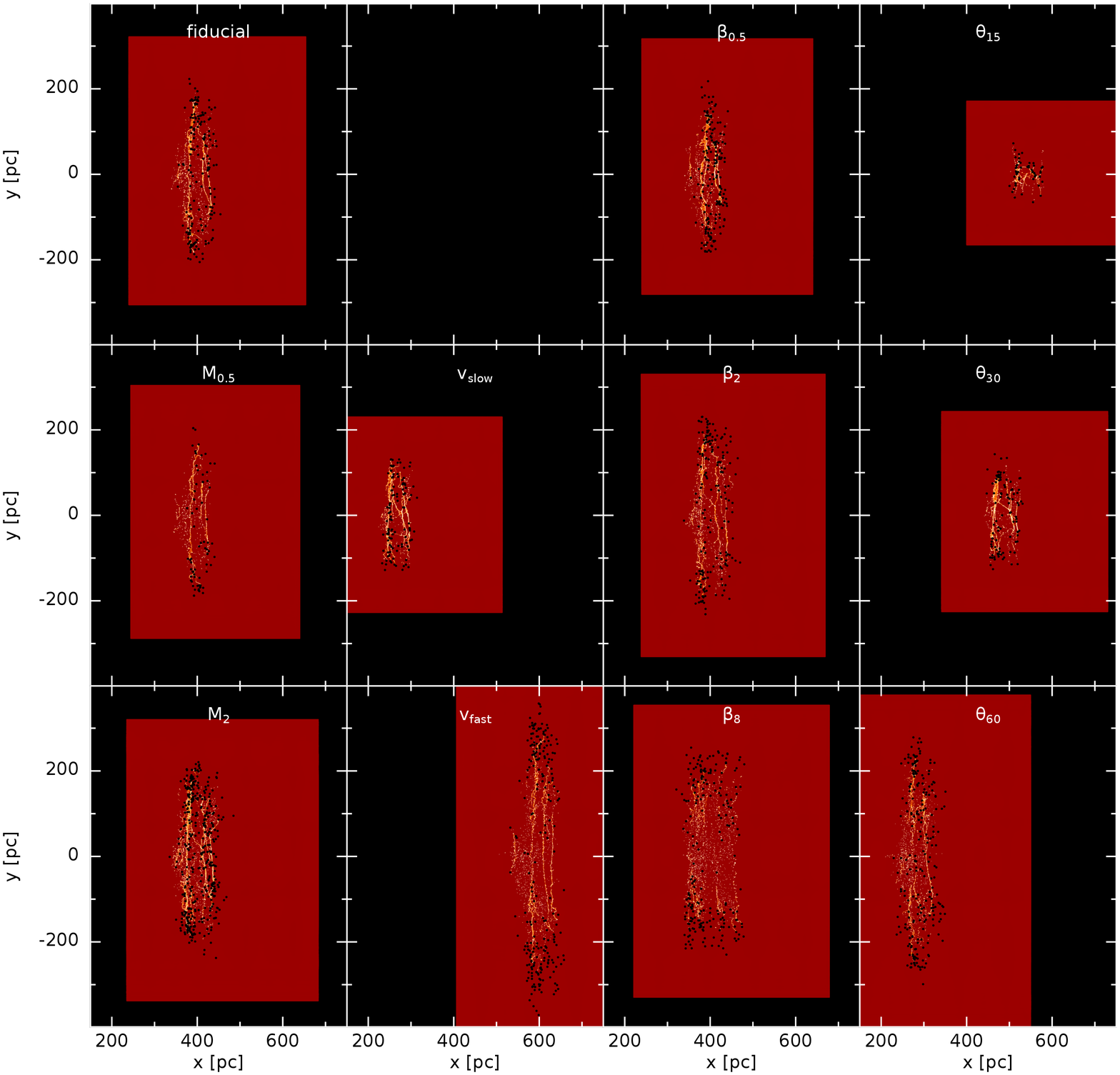}
\includegraphics[width=0.75\textwidth]{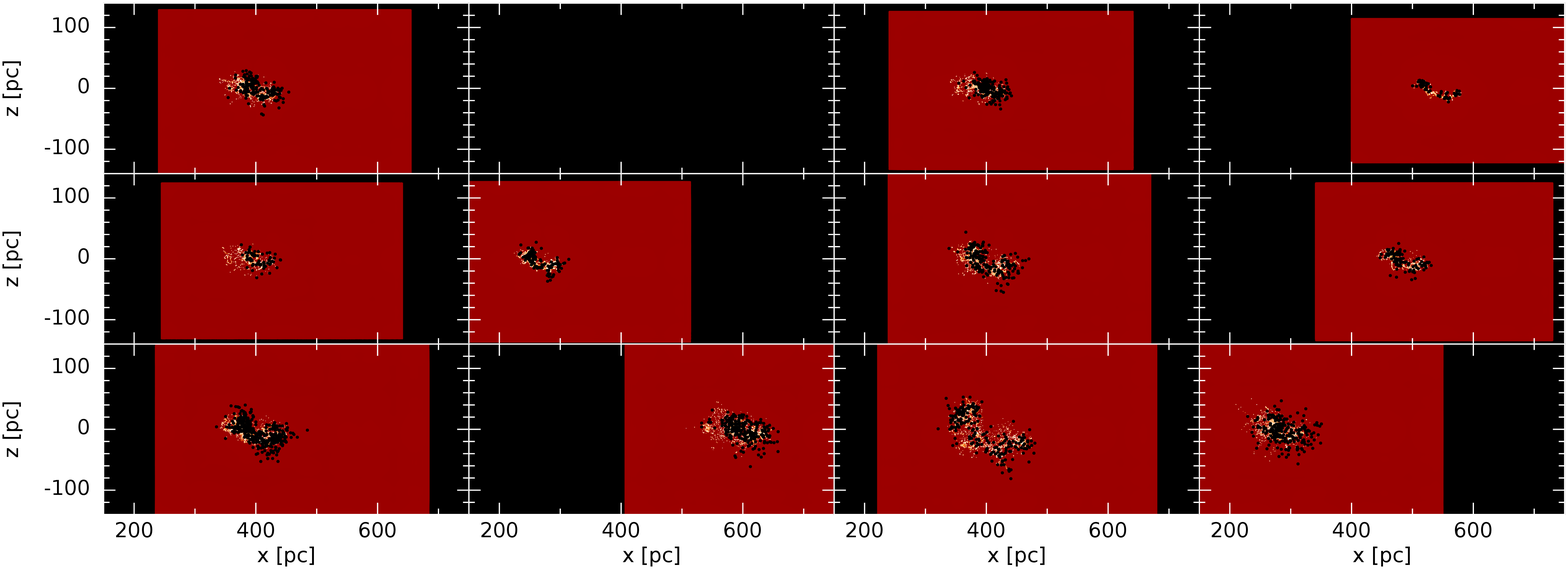}
\includegraphics[width=0.76\textwidth]{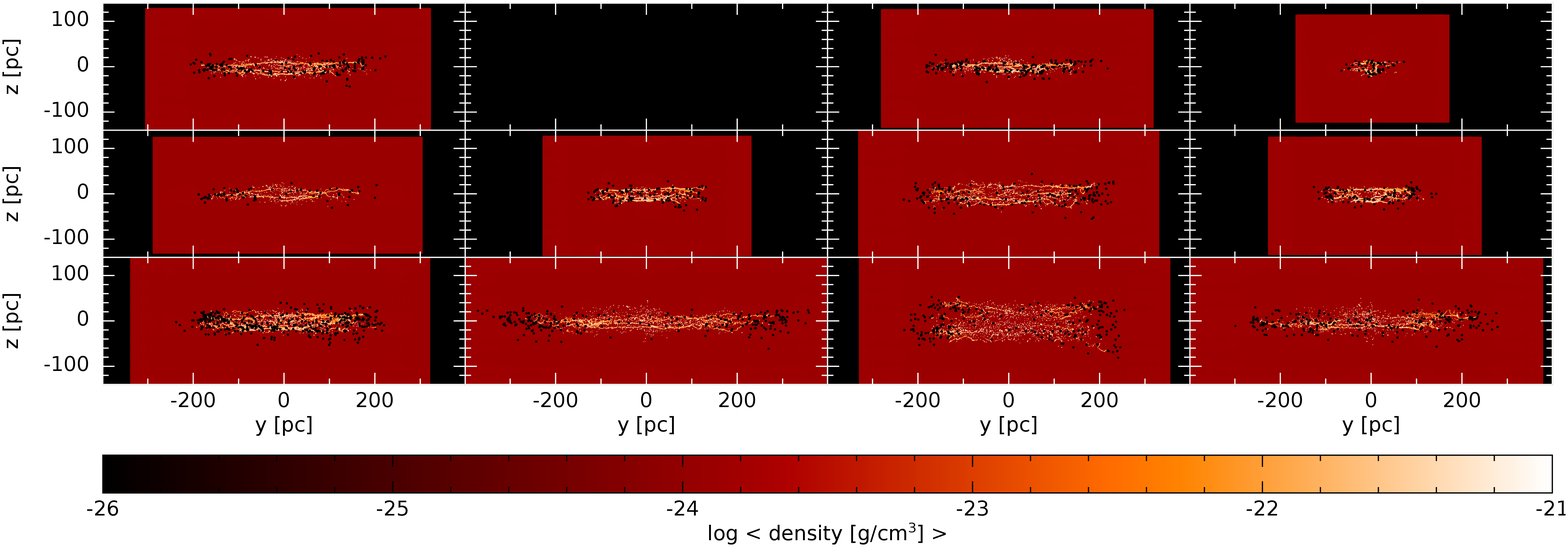}
\caption{Gas density of our suite of simulations after 25~Myr of evolution.  As in \figref{fig:fid:evol}, black represents the region outside of our computational domain, the black dots represent sink particles with a radius 10x that of the sink radius; see \figref{fig:suite:ns} for a version without sinks .  Names are listed in the top panel only for clarity.  The remnant size (in the $y$-direction) and its final location is primarily affected by the initial \vx{} and \vy{} of the cloud.  The level of initial internal turbulence only reasonably affects the remnant if it is high enough, as in \mmachvhigh{}.  The dispersion in the $x$- and $z$-directions is a result of the internal turbulence, both seeded and generated from the collision.} 
\label{fig:suite}
\end{figure*} 

\subsubsection{Effect of the initial conditions}

\figref{fig:suite} shows the remnants after 25~Myr of evolution for each model in our suite of simulations; a version excluding sink particles is shown in \figref{fig:suite:ns} in \appref{app:figs}.

Qualitatively, each remnant is similar comprising of clumps and filaments, where there are generally 2-3 primary filaments that extend in the $y$-direction.  As with \fid{}, the filaments are generally in the $y$-direction, but a small filament in the $-xy$-direction exists in three models (\mslow{}, \mtwo{}, and \manglow{}).  This structure is born after the initial collision and gets stretched between the leading and trailing filaments as they slowly separate and contract.    The exception is \mmachvhigh{}, our most turbulent model, in which filaments never form.  During the collision in this model, the remnant immediately fragments into clumps, which collapse and become more defined as the remnant evolves.  This is consistent with \citet{TanvirDale2020,TanvirDale2021} who found that at least one cloud must be bound for filaments to form.  

The size and position of the remnant varies with the initial conditions.  Naturally, those with slower (\mslow{} and \manghigh{}) or faster (\mfast{}, \mangvlow{}, and \manglow{}) initial \vx{} end at smaller or larger $x$, respectively; as in the fiducial model, the final position in the $x$-direction is not influenced by the background medium.  Similarly, those with slower (\mslow{}, \mangvlow{}, and \manglow{}) or faster (\mfast{} and \manghigh{}) initial \vy{} are less or more extended in the $y$-direction, respectively.  Therefore, in terms of final $x$-position and $y$-extension, the initial impact angle plays the largest role.  

The slight $L$-shape visible in the $xz$-plane (middle panels of \figsref{fig:suite}{fig:suite:ns}) is a result of the initial turbulent velocity field and not dependent on the evolutionary dynamics.  As such, it is more prominent in \mmachvhigh{} than the remaining models.

For each model, there are more sinks associated with the trailing filament than the leading filament, or correlated clumps in the case of \mmachvhigh{}.  This is reasonable given that the trailing filament is created later during the evolution when the bulk of the clouds are colliding.  The sink distribution along the filaments is dependent on the model, with the sinks being clustered towards the end of the filaments in \fid{}, \mhalf{}, \mfast{}, \mmachhigh{}, \mmachvhigh{}, and \manghigh{}; in these models, the sinks that are clustered towards the end of the filaments tend to have intermediate ages of \sm5-18~Myr, while the older sinks ($\gtrsim18$~Myr) are generally distributed throughout the filaments.  In the remaining models, the sinks are relatively evenly distributed along the filaments.  In all models, the sink particles are reasonably confined to the $z\approx0$ plane.

\subsection{Gas dynamics}
\label{sec:gd}

\subsubsection{Gas mass evolution}
\label{sec:gd:me}

\begin{figure} 
\centering
\includegraphics[width=0.95\columnwidth]{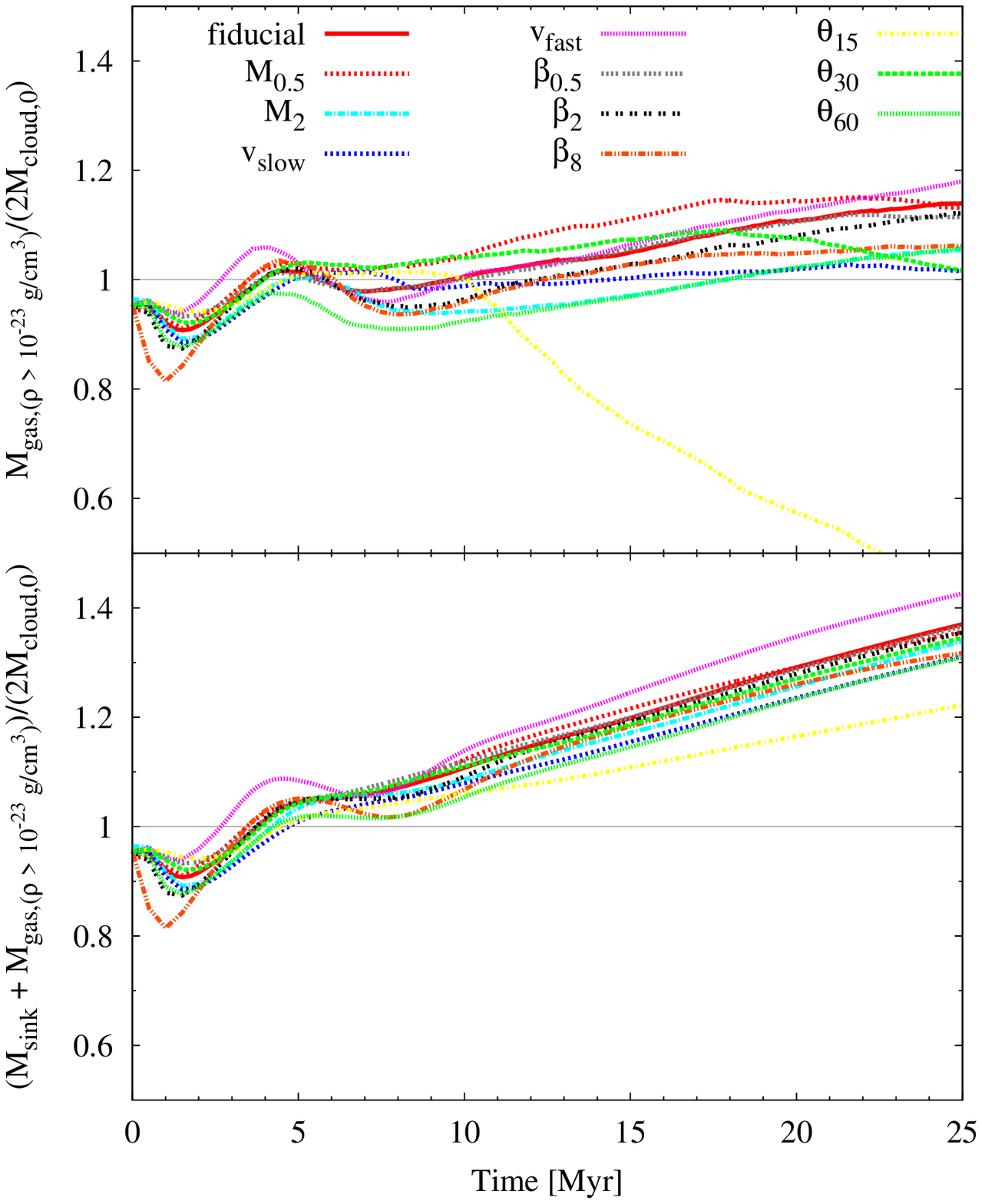}
\caption{Evolution of the quantity of dense gas mass (top) and combined dense gas mass and sink mass (bottom) as a fraction of the total initial cloud masses. We define `dense gas' to be \rhoge{-23},  which is slightly lower than the initial cloud density of $\rho_\text{cloud,0} = 1.2\times 10^{-23}$~\gpercc{}.  However, SPH smoothing generates a smooth transition in density between the cloud and the medium, making this value somewhat arbitrary; this smooth transition also accounts for the initial ratio of $< 1$.  The gas ratio (top) at 25~Myr for \mangvlow{} is 0.43.
As the remnants move through the background medium, they shock gas to higher densities and sweep it up into the filaments and clumps, accounting for the increase in dense gas; the exception is \mangvlow{}, which has a small cross-section due to the small impact angle, yielding a small reservoir from which is can accrete (top).  Both dense and non-dense gas is accreted onto the sinks, accounting for the increase in gas+sink mass for all models (bottom).}
\label{fig:GasMassRatio}
\end{figure}

The evolution of the total mass of dense gas (which we define as gas with $\rho_\text{dense} \ge  10^{-23}$~\gpercc{}) is shown in the top panel of \figref{fig:GasMassRatio}.
Due to the initial numerical smoothing at the edge of the clouds, the initial quantity of dense gas is slightly lower than the defined initial mass of the clouds.  As the evolution begins, there is a slight decrease in total dense gas mass ($t \lesssim 1.5$~Myr).  During this time, the initial turbulence forces some of the outer regions of the cloud into the background medium, decreasing its density and hence the total amount of dense gas; the explains why the decrease is most notable in our most turbulent model, \mmachvhigh{}.   As the clouds independently move through the medium ($t \lesssim 2.2$~Myr), the shear causes a density increase at the trailing edge of the cloud.  By $t \sim 1.5$~Myr, the amount of dense gas created by the shear surpasses the amount lost due to turbulent motions, and the total amount of dense gas begins to increase.  After the collision, there is a notable decrease in gas mass as the dense gas is converted into and/or accreted onto stars (see \secref{sec:sd}).  For $t \gtrsim 7$~Myr, the quantity of dense gas increases.  This is from both the warm medium surrounding the clumps cooling and condensing onto the clumps \citep[as in, e.g.,][]{KoyamaInutsuka2002}, and the remnants sweeping up background gas that has a slower velocity than itself\footnote{This argument is component-wise, not net velocity.}.  In both scenarios, our relatively high background density ($\rho_\text{bkg} \approx 0.11\rho_\text{cloud,0}$) plays a role in the evolution of the clouds and the dense gas, and our remnants continually replenish their reservoir of dense gas.  

By 25~Myr, all models except \mangvlow{} have more dense gas than in the original clouds.  Aside from increasing the initial velocity (\mfast{}), modifying any property away from the fiducial model decreases the final amount of dense gas, although this decrease is small for \mhalf{}, \mmachlow{}, and \mmachhigh{}.  The total amount of dense gas in \mangvlow{} decreases for $t \gtrsim 10$~Myr, since these remnants (filaments and cores) have the smallest cross section of all the models yielding a reduced interaction with the background gas.

\subsubsection{Gas mass distribution}
\label{sec:gd:dis}
\begin{figure} 
\centering
\includegraphics[width=0.85\columnwidth]{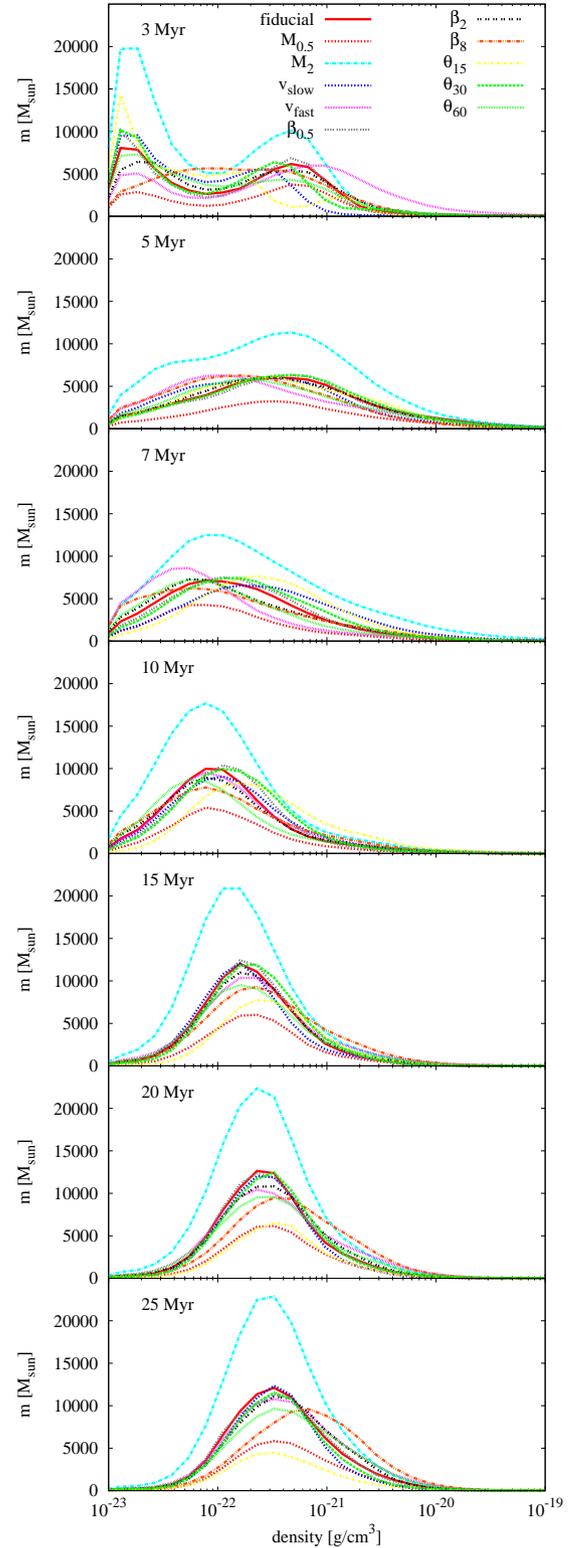}
\caption{Distribution of the gas densities at seven times for \rhoge{-23};  recall $\rho_\text{cloud,0} = 1.2\times 10^{-23}$~\gpercc{}, so we are excluding the background gas in this plot.  The distribution at 3~Myr is defined by the isolated clouds moving through the medium, while the distribution at 5~Myr is defined by the collision.  The shift in the distributions at 7~Myr result from the high-density gas being converted into stars.  By 10~Myr, the distributions have generally settled, but slowly shift to higher densities as the dense gas accretes gas from the background.} 
\label{fig:gasIMF}
\end{figure}

To complement our analysis of the evolution of the total gas mass (top panel of \figref{fig:GasMassRatio}), \figref{fig:gasIMF} shows the gas density distribution function for seven times.  
Just prior to the collision (3~Myr; top panel), the gas distribution is generally bi-modal.  Both of these peaks are in the cool gas, thus does not represent the two-phase medium discussed in the introduction; there is a third peak at lower densities than plotted that represents the warm, low-density background medium.  The two peaks plotted represent the different behaviours of the leading and trailing edges of the pre-collision clouds as they interact with the background (as discussed in \secref{sec:ge}).  The exception to this bi-modality is \mmachvhigh{}, which has uni-modal distribution since the internal turbulence has already generated substructure within each cloud.

At 5~Myr, there is a temporally local maximum of dense gas (top panel of \figref{fig:GasMassRatio}), and the gas density distribution is now more Gaussian (second panel of \figref{fig:gasIMF}).  At this time, the majority of the gas initially in the clouds is involved in the collision, and much of the internal substructure created before the collisions has been washed out; this results in the Gaussian-like distribution with mean density of $\bar{\rho} \gtrsim \rho_\text{cloud,0}$.  

As the collision proceeds over the next few Myr, there is a burst of star formation (\secref{sec:sd}), and much of the high density gas is converted into or accreted onto the sink particles.  This conversion/accretion accounts for the decrease in the high-density tail, and the mean density decreases to \sm10$\rho_\text{cloud,0}$ by 10~Myr.  The different initial conditions yield different intermediary distributions at 7~Myr, but the distributions are qualitatively similar by 10~Myr.  The two outliers are \mhalf{} and \mtwo{}, which is simply a result of their lower and higher initial masses, respectively.  

Once the remnant begins to expand after the collision ($t > 10$~Myr), the entire gas distribution shifts to higher densities as the clumps and filaments accrete the background gas.  The initial turbulent energy of \mmachvhigh{} yield remnant clumps rather than filaments  \citep[recall][]{TanvirDale2020,TanvirDale2021}, which are able to collapse to higher densities than an extended filament, resulting in even higher gas densities than the remaining models.  The lack of gas accretion in \mangvlow{} accounts for a decrease in gas mass at all densities; despite this decrease, the evolution of its distribution follows the same general evolution as the remaining models.  

\subsubsection{Gas velocity function}
\label{sec:gd:v}
\begin{figure*} 
\centering
\includegraphics[width=\textwidth]{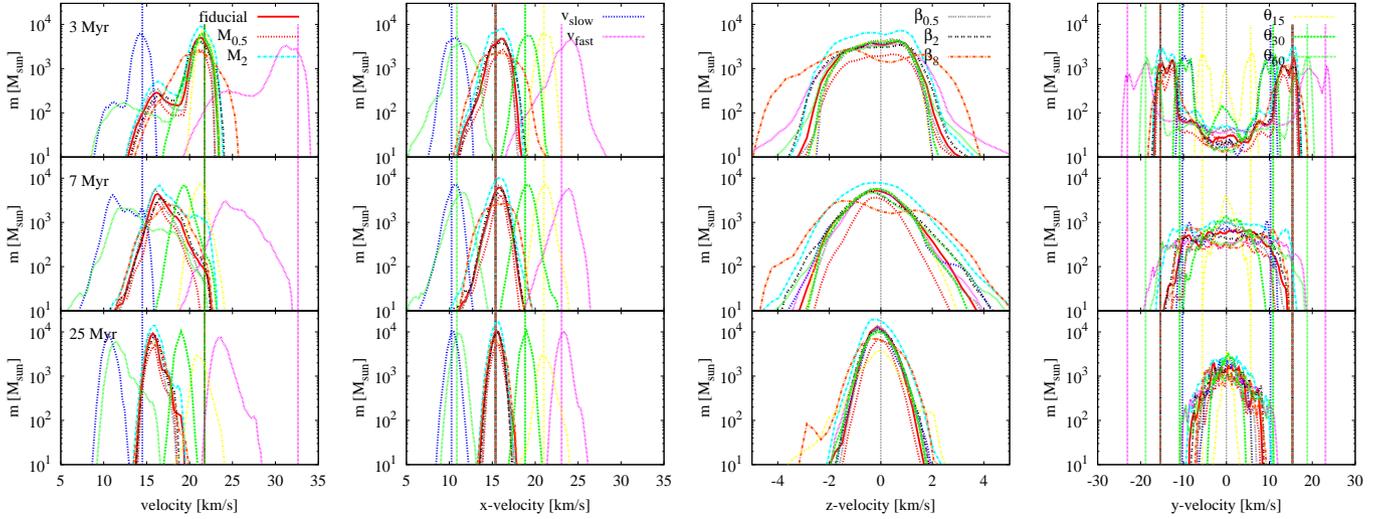}
\caption{Distribution of the gas velocities at 3, 7, and 25~Myr for \rhoge{-23}.  The horizontal scale is different for nearly every column to best highlight the distribution.  The vertical lines of the same colour represent the initial velocity of corresponding model.  The velocity in the $x$-direction is tightly centred on its initial velocity, while the $y$-component decreases with time; the decrease in the $y$-component leads to the overall decrease in velocity from its initial velocity.  The distribution width decreases with time as the turbulence (both the initial turbulence and the turbulence caused by the collision) decays.}
\label{fig:gasVelocites}
\end{figure*}

\figref{fig:gasVelocites} shows the velocity distribution of dense gas at 3, 7, and 25~Myr.  
The clouds first impact at 2.2~Myr, therefore the top row represents the velocity structure in the very early stages of the cloud collisions, however, it is more representative of the pre-collision distributions.  

\begin{figure} 
\centering
\includegraphics[width=\columnwidth]{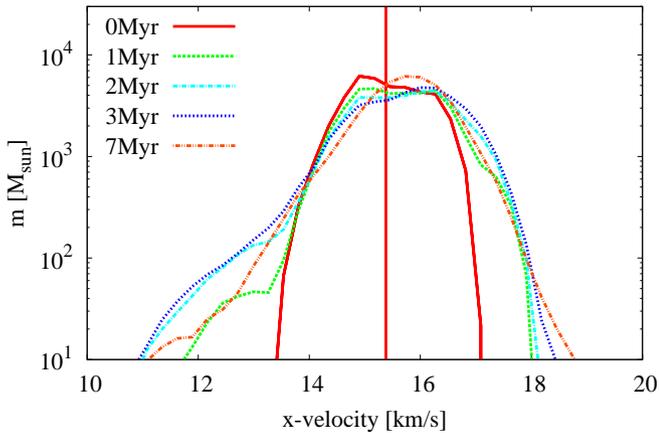}
\caption{Distribution of the \vx{} gas velocity at 0-7~Myr for \fid{}.  The initial velocity peak (0~Myr) is slightly offset due to the initial turbulent velocity.  The mass of gas with $v_\text{x} \sim 16$~\kms{} is similar at all five times.  The turbulence increases the width of the distribution and shear generates the low-velocity tail resulting in a decrease of the actual peak and the apparent shift in the peak to higher velocities.  By 7~Myr, the collision is exciting the gas velocity and pushing it to higher velocities to create the new peak closer to \vxi{}.}
\label{fig:gasVelocites:fid}
\end{figure}

Initially, the distribution of \vx{} is distributed about \vxi{}.  As the system evolves, there are many processes (namely turbulence and shear) that decrease the gas velocity, and these predominantly act on the slower gas.  Therefore, at 3~Myr, the \vx{} distribution is peaked at or slightly higher than \vxi{} (most noticeable in \mfast{}).  This does not represent a shift in the distribution, but rather that the faster gas has not been slowed down. For clarity, \figref{fig:gasVelocites:fid} shows the \vx{} distribution between 0 and 7~Myr for \fid{}, which shows that between these times, the amount of gas with $v_\text{x} \sim 16$~\kms{} is approximately constant, and the apparent shift in the peak is due to the physical processes preferentially slowing the slower velocity gas. 

Although some of the gas retains its initial \vy{}, much of it has slowed down; this also contributes to the low-velocity tail in the total velocity.  This is simply a result of the clouds moving into the background gas that has no initial \vy{} and slowing down due to ram pressure.  At this time, the width of the distribution of all three components is a result of the initial internal turbulence; the higher initial turbulence (\mmachvhigh{}) or faster initial velocity (\mfast{}) slightly broaden this distribution.

Near the end of the collision at 7~Myr (middle row of \figref{fig:gasVelocites}), the final qualitative shape of the distribution has taken form.  The distributions are less broad in all components, and the initial bi-modal distribution in \vy{} has been completely washed out by the collision.  The peak of \vx{} has shifted back closer to \vxi{} as the high-velocity gas is slowing as it interacts with the background, but the peak of the total velocity has decreased primarily due to the decrease in \vy{}.  

As the gas evolves to 25~Myr, the distributions continue to narrow in all components and in the total velocity.  By this time, the remnant has fragmented into cores and filaments.  The cross section of these objects is small enough such that they are strongly influenced by the background gas as they interact with it.  Hence, the background promotes these filaments and clumps to move with \vxi{} (where the peak of the distribution is now centred and has been since 15-20~Myr, depending on the model) and has removed the `high-velocity' peak visible at earlier times; further the background gas hinders all motion in the $y$- and $z$-directions.  There is still some expansion in these directions (recall \figref{fig:fid:evol}), however it is slowing with time will continue to do so as the remnants evolve.  

In agreement with previous studies \citepeg{KoyamaInutsuka2002,Vazquezsemadeni+2007}, the background medium plays an important role when modelling a two-phase medium.

\subsection{Star formation and evolution}
\label{sec:sd}

To permit the long-term evolution of our simulations, we replace dense, collapsing regions with sink particles of radius 0.25~pc; therefore, these particles represent star forming clusters\footnote{Until further notice, we interchange star formation, star cluster formation, and sink formation.}.  Star cluster formation beings only after the clouds collide, despite the clouds being initially turbulent;  this was by design of our fiducial model.  During the collision, there is a bust of star cluster formation (recall \figref{fig:fid:evol}), and then star clusters are continually formed for the duration of the simulation, albeit at a slower rate. 

The bottom panel of \figref{fig:GasMassRatio} shows the total dense gas plus sink mass as a function of time for each model.  After the collision, there is a steady increase in total mass, and by 25~Myr, there is 20-40 per cent more dense+sink material than initially in the clouds.  \mfast{} has the largest total mass while \mangvlow{} has the least; these models have the largest and smallest extent in the $y$-direction, respectively, thus have accreted the most and least amount of gas due to the differential \vy{} between the clumps and the background material.  This again shows that there is a reasonable amount of background gas that is accreting onto the dense objects, whether it be a gas clump/filament or a sink particle.   Therefore, if the background gas has a comparable density to the clouds, then the background will be dynamically important.

\begin{figure} 
\centering
\includegraphics[width=\columnwidth]{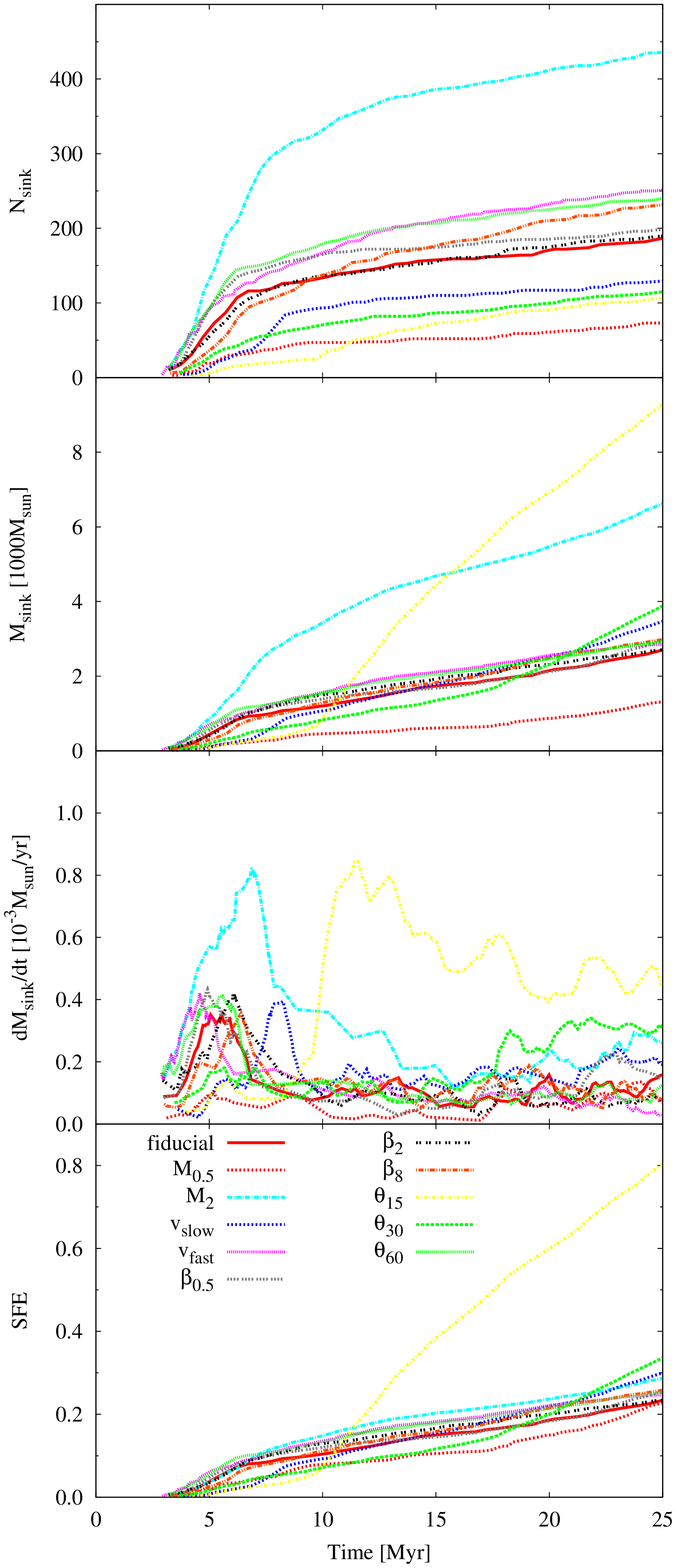}
\caption{From top to bottom: The total number of sink particles, the total mass in stars (i.e., sinks), the total mass accretion rate onto sinks (moving averaged over 1~Myr), and the star formation efficiency (i.e., the total mass in stars divided by the initial mass of both clouds).  Note that the mass accretion rate includes both gas turned into sink particles and gas accreted onto existing sinks.  The total number of sinks depends primarily on the initial mass of the cloud and the initial \vy{}.  Model \mangvlow{} is an outlier with respect to the other models since the remnant gas from the collision is well-confined to $y \approx 0$, permitting the relatively few stars that form to accrete a disproportionally large amount of gas.  }
\label{fig:sinkNME}
\end{figure}

\figref{fig:sinkNME} shows the total number of sinks, total mass in sinks, the sink formation rate, and the star formation efficiency as a function of time for each model in our suite.
All models follow similar trends.  The first star cluster forms between \sm3-4~Myr; the small spread of when the first sink forms is not significant, thus we conclude that none of our initial conditions have the ability to delay the onset of star formation.  By 25~Myr, there are generally \sm100-250 sinks with a total mass of \sm2000-3000~\Msun{} in stars.  

Unsurprisingly, \mtwo{} produces just over twice the number of stars as the fiducial model given it has twice the initial mass in the clouds (top panel).  This is primarily from the initial star burst, which produces \sm300 and \sm100 stars for \mtwo{} and \fid{}, respectively; see also the third panel, which shows that the initial star burst is both higher and longer for \mtwo{} than \fid{}.  This longer star burst is a result of the initial clouds in \mtwo{} having a longer semi-major axis thus the collision occurs over a slightly longer time.  Also as expected, there is more total mass in stars in \mtwo{} than \fid{}, however, as a fraction of the initial cloud mass, this is only \sm20-40 per cent times greater for \mtwo{} after the end of the initial star burst (bottom panel).  Therefore, doubling the initial mass of the initial clouds approximately scales up the stellar cluster properties by a factor two.  

Similarly, \mhalf{} produces slightly less than half the stars as the fiducial model with about half of the total stellar mass.  This model maintains a relatively constant star formation rate and does not include a starburst during the initial collision.    The star formation efficiency (bottom panel) is lower in \mhalf{} than \fid{} until \sm24~Myr, when a small star burst occurs, yielding a similar SFE at 25~Myr between the two models.  

For the models with the same initial cloud mass (i.e., all models except \mhalf{} and \mtwo{}), the total number of sinks at 25~Myr varies between 100-250.  Small changes to the initial level of turbulence has minimal effect on the total number of stars (i.e., \mmachlow{} and \mmachhigh{}),  although increasing it as in \mmachvhigh{} increases the final number of stars by \sm25 per cent compared to \fid{}.

Of our parameters, the initial \vy{} has the largest impact on the number of star clusters that form.  Deceasing \vy{} as in \mslow{}, \mangvlow{}, and \manglow{} decreases the number of sinks compared to \fid{}, with the decreasing order for $t < 7.5$~Myr shown in the top panel of \fig{fig:sinkNME} corresponding to the decrease in initial \vy{}; at \sm7.5~Myr, \mslow{} undergoes a brief star burst at which time this trend breaks.  This delayed star burst in \mslow{} is somewhat artificial, since we set initialised the clouds' locations such that the cloud impact time was constant across all models, but with \mslow{}, additional time is required for the bulk of the mass to collide and reach star forming densities.  Therefore, it is likely that there are no physical differences amongst our initial conditions that can delay a star burst.

Increasing \vy{} as in \mfast{} and \manghigh{} increases the number of sinks compared to \fid{}, again with the number of sinks increasing with increasing \vy{}.  The faster initial velocity in \mfast{} permits the bulk of the cloud mass to collide and reach star forming densities earlier than the other models, however, this earlier star burst is not significantly earlier than any of the other models.  This correlation between initial velocity the the number of sinks is consistent with \citet{LiowDobbs2020}.

Since all the gas in any given model is initialised with the same \vx{}, larger \vy{} means that the gas will diverge more quickly after the collision (recall the extent in the $y$-direction as shown in \figref{fig:suite}), resulting in the formation of more, but less massive, sinks \citep[again, in agreement with][]{LiowDobbs2020}.  This is confirmed in the second panel of \figref{fig:sinkNME} where the total stellar mass is similar (except \mangvlow{}) meaning more low-mass stars  in the models with larger \vy{}; see also \secref{sec:imf}.

In terms of total mass and the mass accretion rate (second and third panels of \figref{fig:sinkNME}, respectively), the clear outliers are \mhalf{} and \mtwo{} (as discussed above) and \mangvlow{}.    \mangvlow{} consistently has the lowest number of stars, however, by $t \sim 10$~Myr it has the most total mass in stars of all the models with the same initial cloud mass, and by $t \sim 15$~Myr it has the most total mass in stars of all our models.  Therefore, these stars in \mangvlow{} are rapidly accreting gas and becoming very massive (see also \secref{sec:imf}).  By about 10~Myr, \vy{}\sm0, meaning that the dense gas from the collision has nearly stopped dispersing.  Therefore, a massive and localised reservoir of gas remains from which the sinks very efficiently and continuously accrete -- this reservoir is unique to this model.  Given the relatively small number of stars, some form of competitive accretion likely promotes the higher accretion rate since there are fewer stars competing for the same material.  From the high SFE of \mangvlow{} (bottom panel), it is reasonable to expect that these stars have also accreted a reasonable amount of material that was initially in the background medium. 

\subsubsection{Sink mass function}
\label{sec:imf}
\begin{figure} 
\centering
\includegraphics[width=\columnwidth]{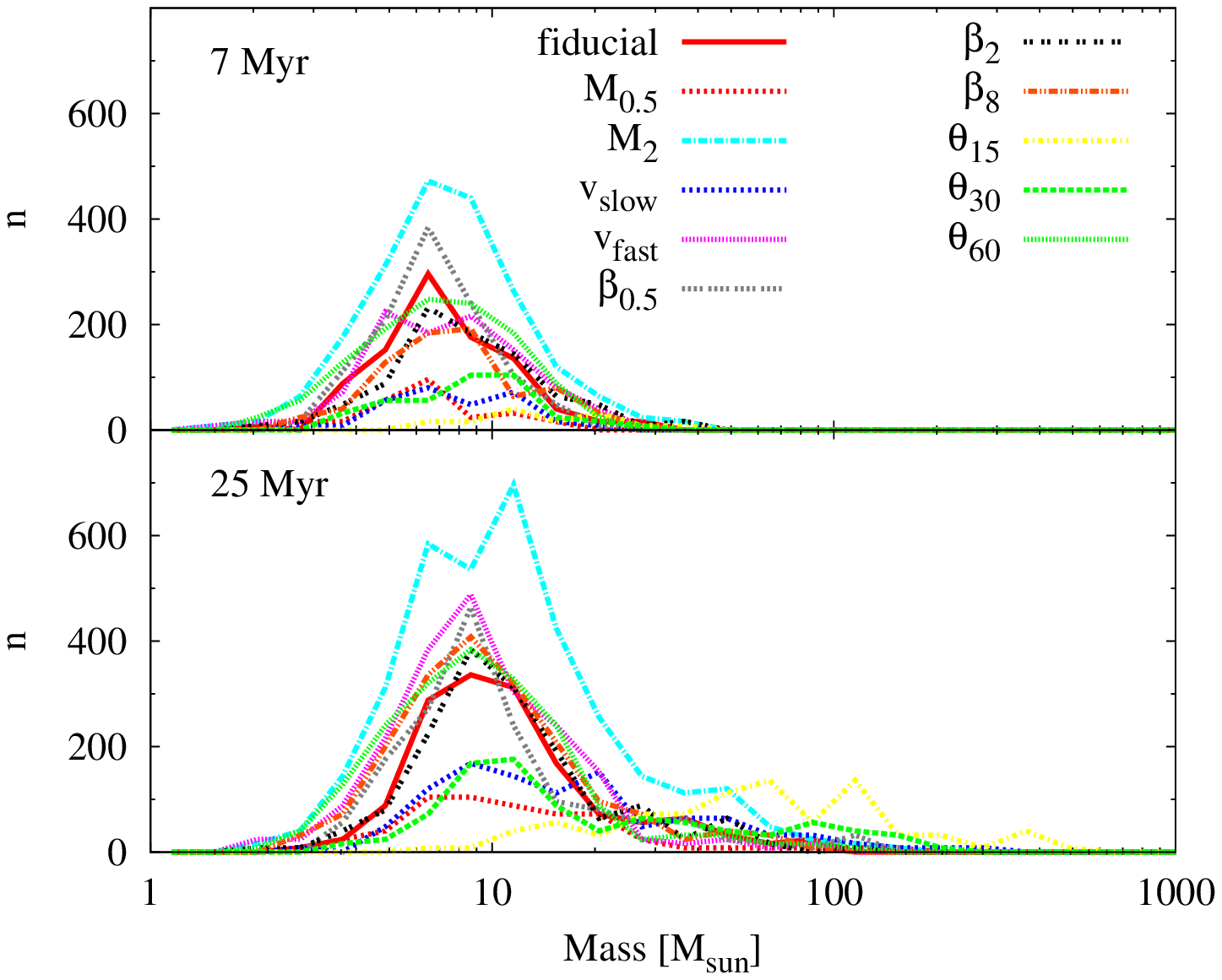}
\includegraphics[width=\columnwidth]{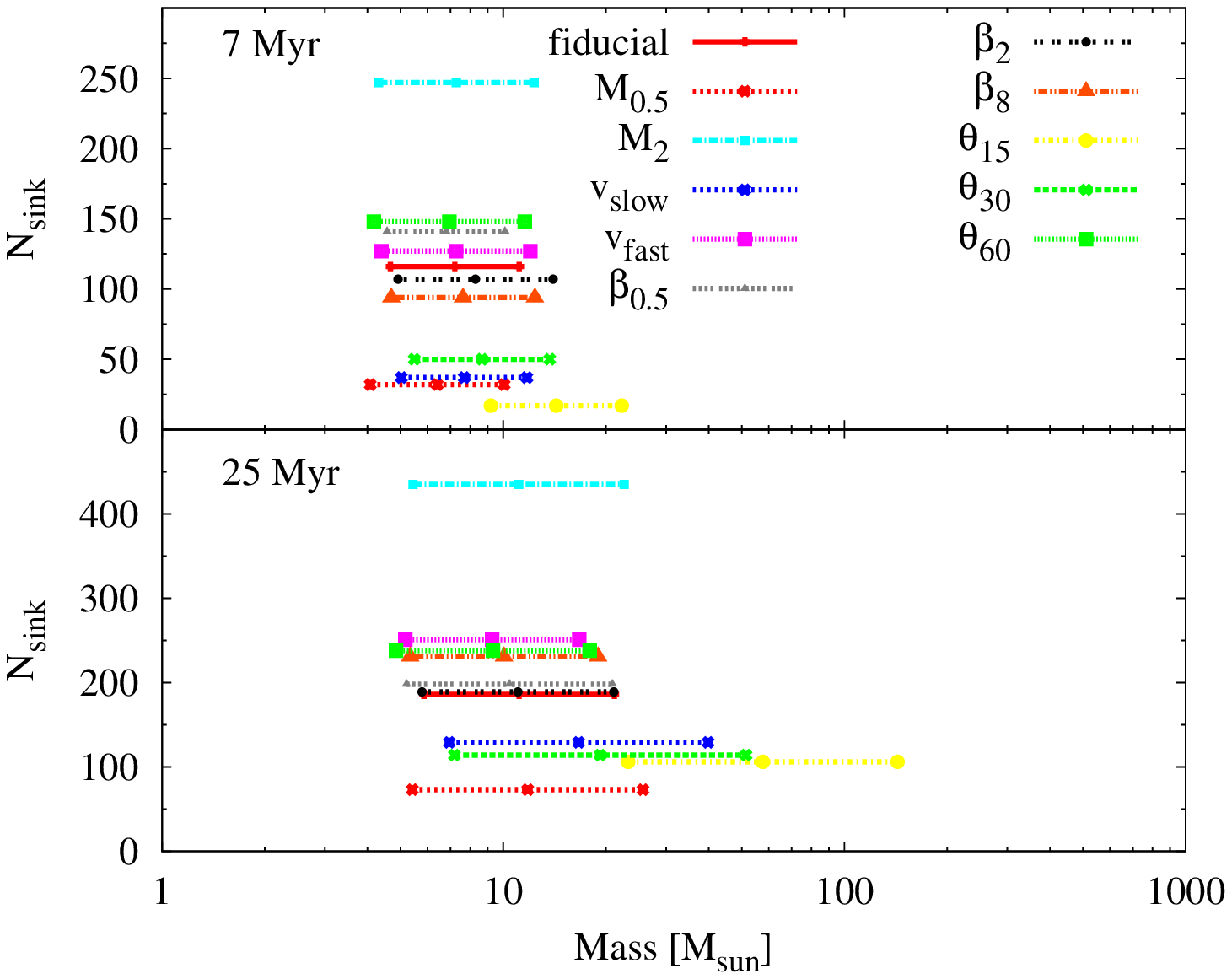}
\caption{\emph{Top panel}: Sink mass functions at two different epochs; the bins are evenly spaced in log-space.  \textit{Bottom panel}: Each line covers the average sink mass for each model plus/minus one standard deviation; the average and standard deviation were calculated in log-space for comparison with the top panel.  Both panels show that the average sink mass increases and the mass distribution broadens with time.  The resulting average and standard deviation are sensitive to slow initial collisions in \vyi{} where the remnant gas remains relatively confined.} 
\label{fig:sinkSMF}
\end{figure}

The sink particles in our simulations have radii $r_\text{sink} = 0.25$~pc, thus represent star forming regions rather than individual stars.  The minimum resolved sink mass is \sm0.33~\Msun{}, although very few sinks have this mass, even at birth.  While the low-mass sinks may represent just a few stars, high mass sinks clearly represent entire clusters that may include a few high-mass stars; given that these sinks represent clusters, we cannot comment on the formation mechanism of individual high-mass stars.  The top panel of \figref{fig:sinkSMF} shows the sink mass functions at 7 and 25~Myr, and the bottom panel shows the average sink mass $\bar{m}$, and $\bar{m} \pm \sigma$.

By 7~Myr, the primary star formation epoch is complete (recall top panels of \figref{fig:sinkNME}).   Although the number of sinks varies amongst models, all models have similar averages ($\bar{m} \approx 7-9$~\Msun{}) and standard deviations (calculated in log-space), indicating that our initial conditions have a minimal role on the initial sink mass distribution.  The exception is \mangvlow{}, whose average sink mass is $\bar{m} = 14.3$~\Msun{}; as discussed above, this is a result of the gas being well confined to the $y\approx0$ plane to provide a larger gas reservoir than in the other models.  

At 25~Myr, each mass distribution has a reasonably high-mass tail, with sinks growing up to \sm300~\Msun{}, which is \sm5 per cent the initial mass of the fiducial cloud.  When excluding \mslow{}, \manglow{}, and \mangvlow{}, the majority of the sinks in each model tend to have $m \lesssim 25$~\Msun{}, with an average mass of $\bar{m} \approx 9-12$~\Msun{}; these averages are only \sm25 per cent higher than the averages at 7~Myr.   As at 7~Myr, despite the different total number of sinks, each model has a similar average and standard deviation.  

The three exceptions listed above are those that have initially slower \vy{} than \fid{}.  As discussed in \secref{sec:sd}, this slower \vy{} permits each sink to maintain a larger gas reservoir from which it can accrete.  In the case of \mangvlow{}, this permits several massive sinks ($50 \lesssim m/$\Msun{}$ \lesssim 200$) to form.

The mass distribution of the models with an initial \vy{} faster than \fid{} have no significant distinction from the remaining non-slow models.  Therefore, the average sink mass is higher and the distribution is broader when clouds collide slowly and the remnant gas is relatively confined.

\subsubsection{Sink velocity function}
\label{sec:sd:v}
When a sink particle forms, its initial velocity is that of its progenitor gas.  As it evolves, it gains the momentum of any gas it accretes, although this becomes relatively less as the sink become more massive.  Given that sinks do not feel gas pressure, they more easily escape a filament than can a gas clump; this can be clearly seen in \figrref{fig:fid:evol}{fig:suite} where the gas clumps are reasonably contained to and near the filaments while many sinks have migrated away.  As discussed above, this migration permits the sinks to accrete pristine gas (note that this pristine gas does not need to cool before being accreted), but also permits a deviation from the gas velocity distribution.  

\begin{figure*} 
\centering
\includegraphics[width=\textwidth]{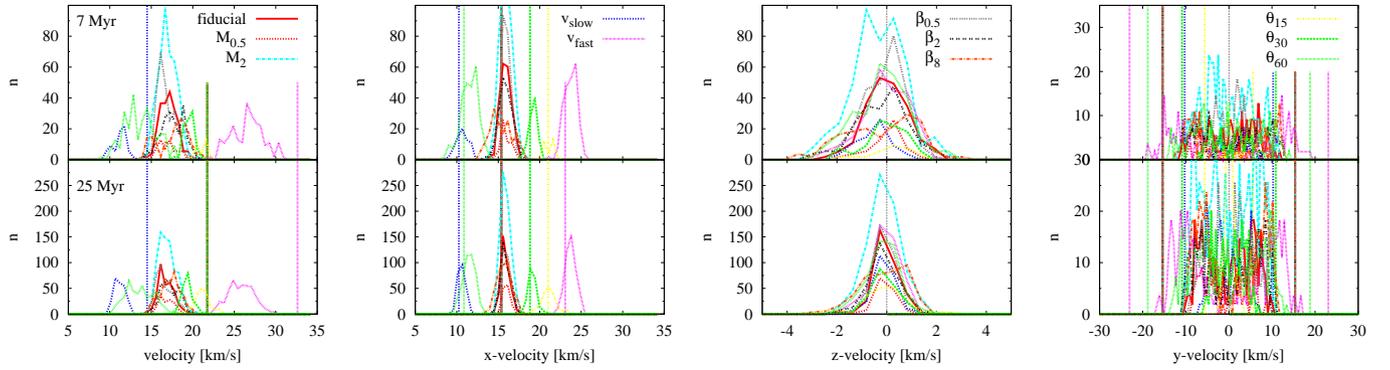}
\caption{Distribution of the sink velocities and their components at 7 and 25~Myr.  The scales are different on nearly every panel to better highlight the distributions.  The vertical lines of the same colour represent the initial velocity of corresponding model.  The horizontal range in each panel is the same as in \figref{fig:gasVelocites} for direct comparison of distribution width; recall that sinks have not formed by 3~Myr.  The sink velocity distribution is generally narrower than the gas velocity distributions, and is generally peaked at slightly faster velocities than the gas.  This is a result of the sinks retaining the velocity of the progenitor gas clump and fewer physical processes are available to decrease the sink velocity.} 
\label{fig:sinkVelocities}
\end{figure*}

\figref{fig:sinkVelocities} shows the distribution of sink velocities for each model in our suite at two different times; we show both the total velocity and the velocity in each component.  
The velocity distribution and its components are broadly similar at both times, although the distribution is smoother at 25~Myr due to the larger number of sinks.  This suggests that sinks well-retain their birth velocity, as described above.  

\begin{figure*} 
\centering
\includegraphics[width=\textwidth]{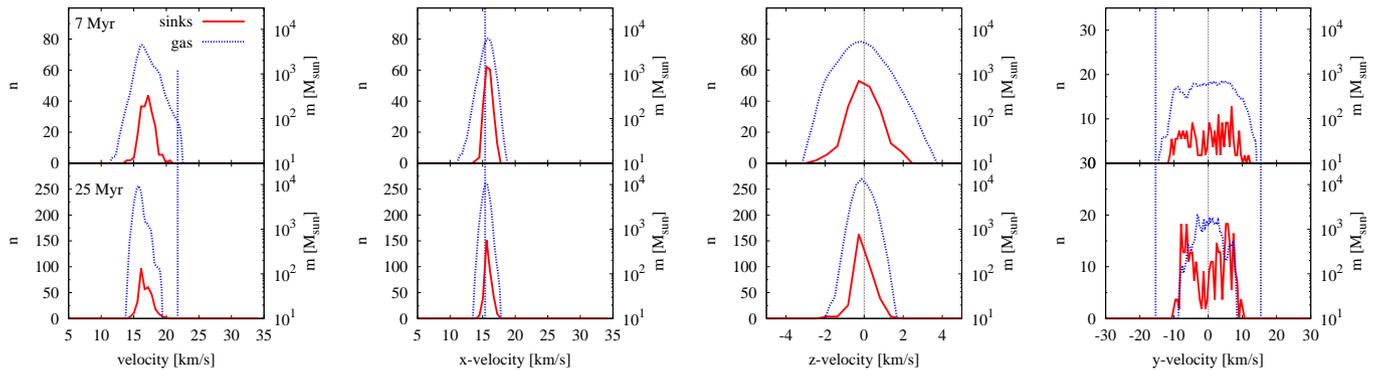}
\caption{Distribution of gas and sink velocities and their components at 7 and 25~Myr for the fiducial model.  The data is as in \figsref{fig:gasVelocites}{fig:sinkVelocities}, but replotted on a single plot for direct comparison.  Except in the $y$-direction, the sinks have a narrower velocity distribution than the gas.} 
\label{fig:Vel:Comp}
\end{figure*}

The sink velocity distribution is notably different than the gas distribution; c.f. \figsref{fig:gasVelocites}{fig:sinkVelocities}, and see \figref{fig:Vel:Comp} for a direct comparison for \fid{}.  In all cases (except \vy{} at late times), the sink distribution is narrower, indicating that the gas on either end of the gas distribution does not collapse to form sinks.  As with the gas, the peak of \vx{} is generally faster than \vxi{}, and in some cases at 7~Myr is faster than the peak of the gas distribution; these latter cases indicate a burst of star formation at earlier times when the dense gas included faster motions.  In most models, unlike the gas velocities, the peaks of the \vx{} distribution does not rapidly shift back to \vxi{}, most notable in \mslow{}.  As star formation proceeds, the stars are formed from gas near the current velocity peak, thus the apparent, but weaker, shift towards \vxi{}.  This is most notable in the models that undergo considerable star formation after 7~Myr, such as \mangvlow{} and \manglow{}.

The distribution in \vy{} is broad for the duration of the simulation for the sinks while it narrows for the gas (readily apparent in the fourth column of \figref{fig:Vel:Comp}, and visualised by the sinks slightly beyond the end of the filaments in \figrref{fig:fid:evol}{fig:suite}).  These sinks continue to move away from the dense regions in the $y$-direction and, with no pressure forces, continue to move unabated.  The gas, however, undergoes a slow decrease in \vy{} due to ram pressure with the background, hence the growing divergence of the \vy{} distributions between the gas and sinks.  If the simulations were to continue beyond 25~Myr, it is likely the velocity distribution of the gas would narrow to \vy \appx 0 (as in \mangvlow{}) while the velocity distribution of the sinks would remain near its current width, and the population of clearly escaping stars would grow.

\subsection{Clumps and filaments}
\label{sec:clumps}

A visual inspection of \figref{fig:suite} shows that the remnants are composed of clumps and filaments at 25~Myr, except \mmachvhigh{} which contains only clumps.  The clumps appear to be within or just beyond the filaments (more clearly seen in \figref{fig:suite:ns}), rather than at hubs formed by the intersection of converging filaments.  Moreover, we do not observe sheets, suggesting that head-on collisions may be required to form sheets.

We identify and analyse the clumps using our clump-finding algorithm that is described in \appref{app:clump}.  The clumps can contain gas and/or sinks, however, 80-90 per cent of all clumps in each simulations are comprised only of gas; see Table~\ref{table:clumps} for the number of clumps in each simulation and the number of clumps of various gas-sink compositions.  This low percentage of hybrid gas-sink clumps is a result of the sink dynamics described in \secref{sec:sd:v} and how the sinks detach from the gas.  In a few cases, there are multiple sinks in a gas-sink clump, and in even fewer cases there are sink-sink clumps without gas.  The total number of clumps per model is roughly correlated with the number of sinks in that model (c.f. \secref{sec:sd}).
\begin{table*}
\centering
\begin{tabular}{l r r r r r r}
\hline
model  & $N_\text{clump}$ & $N_\text{gas only}$ & $N_\text{gas + 1 sink}$ & $N_\text{gas+many sinks}$  & $N_\text{1 sink}$ & $N_\text{many sinks}$  \\
\hline
\fid{}                &   1322  &1149 &   26 &    3  & 141&    3  \\
\mhalf{}           &  706  & 641 &   12    & 2   & 46   &  5   \\
\mtwo{}           & 2213&  1901  &  31 &   20 &  254  &   7  \\
\mslow{}          &   836&   750  &  22   & 11   & 52   &  1  \\
\mfast{}            &   1965 & 1728  &   9  &   2&   220  &   6 \\
\mmachlow{}   &     1108 &  944   & 16  &   9 &  132   &  7   \\
\mmachhigh{} & 1504  &1331  &  10 &    3  & 156  &   4  \\
\mmachvhigh{} &  2150 & 1926 &   21&     2 &  197  &   4  \\
\mangvlow       &       240 &  204 &   24 &   12   &  0  &   0  \\
\manglow{}      &  731 &  673  &  20  &  10  &  26   &  2   \\
\manghigh{}    &   1679&  1449    &12    & 2 &  212    & 4    \\

\hline
\end{tabular}
\caption{The number of clumps per model, categorised by the composition (gas and/or sinks).  Most clumps (80-90 per cent for each model) contain only gas particles. }
\label{table:clumps}
\end{table*}

\begin{figure*} 
\centering
\includegraphics[width=\textwidth]{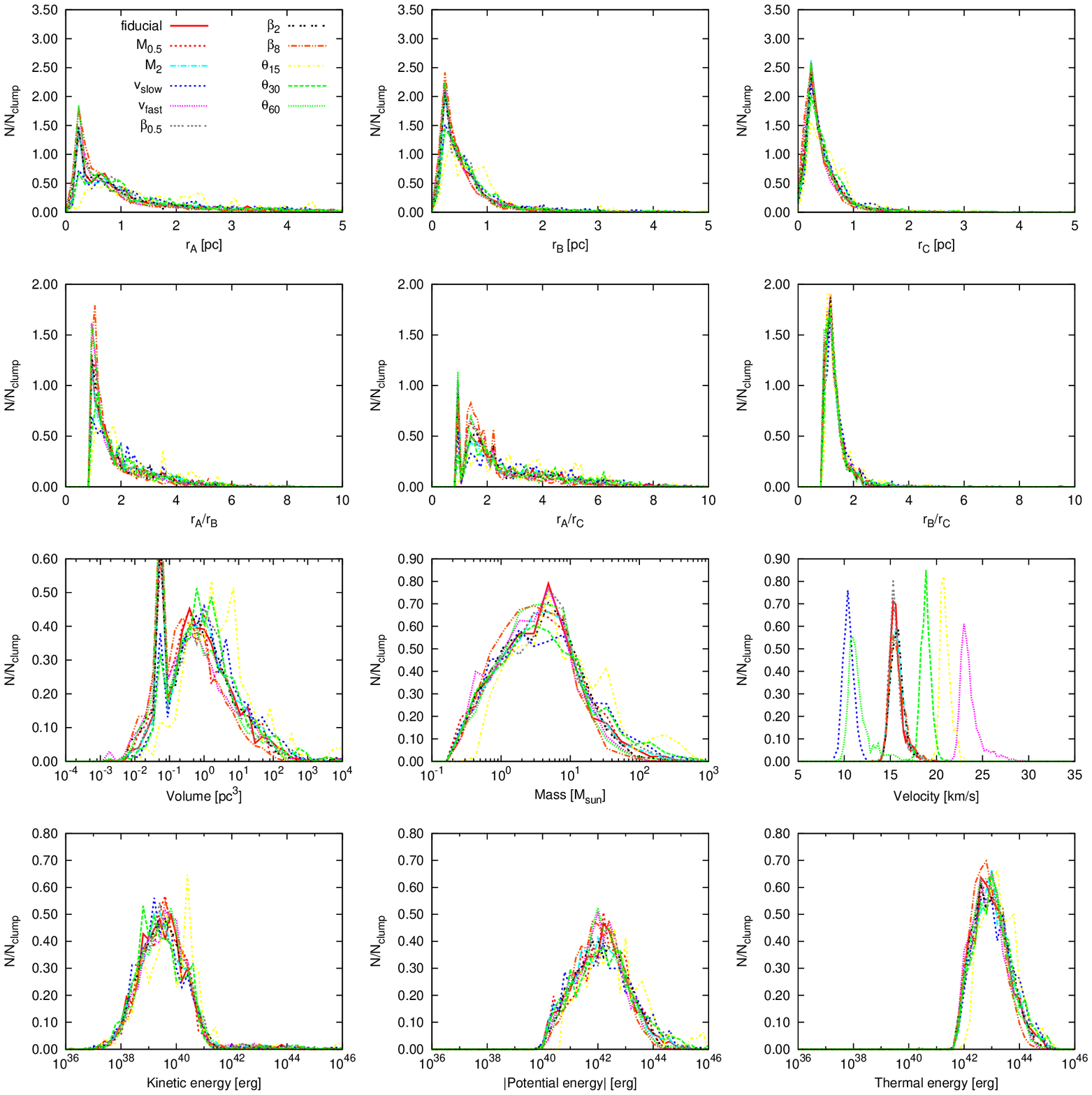}
\caption{Normalised histogram of various clump properties.  Horizontal and vertical axes are different amongst the plots to highlight the curves, but we have tried to maintain consistency amongst similar plots.  The bins are evenly spaced in linear (log) space for the elliptical axes, elliptical axis ratios, and velocity (volume, mass, and energy).  The peak in volume at 0.065~pc$^3$ represents the volume of the sink-only clumps.  The distribution of clump properties is nearly identical for all models, with the exception of the clump velocity.} 
\label{fig:clumpHisto}
\end{figure*}

\figref{fig:clumpHisto} shows 12 histograms of clump properties; each histogram has been normalised by the total number of clumps for better comparison.  
The primary conclusion from this figure is that the normalised clump distribution for each property, except velocity, is very similar for each model.  Therefore, the distribution of clump properties is approximately independent of the initial conditions of the cloud.  

When fitting ellipses to the clumps, the clumps have semi-major axes of $r_\text{A} \lesssim 2.5$~pc, while the remaining two axes are $r_\text{B}, \ r_\text{C} \lesssim 1.5$~pc.  Thus, the bound clumps that we identify are the small dots in \figref{fig:suite}; this indicates that the filaments themselves are not bound structures, although bound clumps may exist within them.  Based upon the axis ratios, the clumps are triaxial but with a preference towards oblate; the spherical clumps are the sink-only clumps, where  $r_\text{A,B,C} \equiv r_\text{sink} = 0.25$~pc.  

The mass of the clumps is generally $m \lesssim 40$~\Msun{}, although there are some more massive clumps in \mangvlow{} which also has a slightly flatter distribution; this is a result of the remnant gas remaining approximately confined to the $y \approx 0$ plane.  The mass range and distribution is similar to that of the sink particles (c.f. \figref{fig:sinkSMF}), although this distribution extends to lower masses since sinks can accrete pristine background gas to increase their mass while these clumps cannot.  This shows that clumps and sinks are simply different realisations of the same objects, lending credit to the sink formation and accretion algorithms.  

The distribution of the properties is nearly identical for all models, indicating that the properties of the clumps is approximately independent of the initial cloud properties.  

\subsection{Resolution}
\label{sec:res}
In numerical simulations, fragmentation is governed by numerical resolution in addition to the included physical processes \citepeg{MeruBate2012,Meyer+2018}.  To test the robustness of our results in the preceding sections, we ran three additional simulations at lower mass resolutions; given the long wall-clock runtime of \fid{}, it was not feasible to perform a simulation at a resolution higher than presented above.  

\begin{figure} 
\centering
\includegraphics[width=\columnwidth]{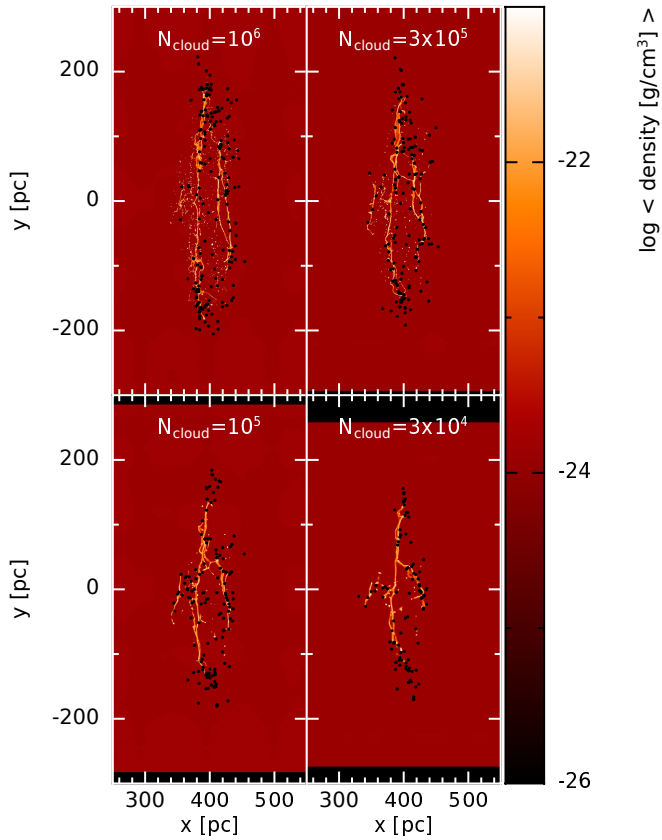}
\caption{Gas density of the fiducial model at four resolutions.  Naturally, there is greater fragmentation and thinner filaments at higher resolutions.   The remnant is more extended in the $y$-direction at higher resolutions, since they are better able to resolve the higher gas and clump velocities.  Convergence has clearly not been reached.} 
\label{fig:fid:resolution}
\end{figure}

\figref{fig:fid:resolution} shows \fid{} at 25~Myr at the four resolutions of $10^6$,  (the fiducial resolution presented above), $3\times 10^5$, $10^5$, and $3\times 10^4$ particles in each cloud, where the latter three models are named \fidm{}, \fidlish{}, and \fidl{}, respectively.  
Although the models are qualitatively similar, there is a clear resolution dependence.  As resolution is increased, there is more fragmentation, and the percentage of gas-only clumps decrease with decreasing resolution; 54 per cent of the clumps in \fidl{} are gas-only compared to 87 per cent in \fid{}.  
The remnant and sink particles become more extended in the $y$-direction at higher resolution since these models are better able to resolve shocks and higher gas velocities and hence sink velocities.  The distribution of sink locations is similar at all four resolutions, although, like the gas, there is a broader distribution of sinks in the $y$-direction at higher resolutions.  

\begin{figure} 
\centering
\includegraphics[width=0.95\columnwidth]{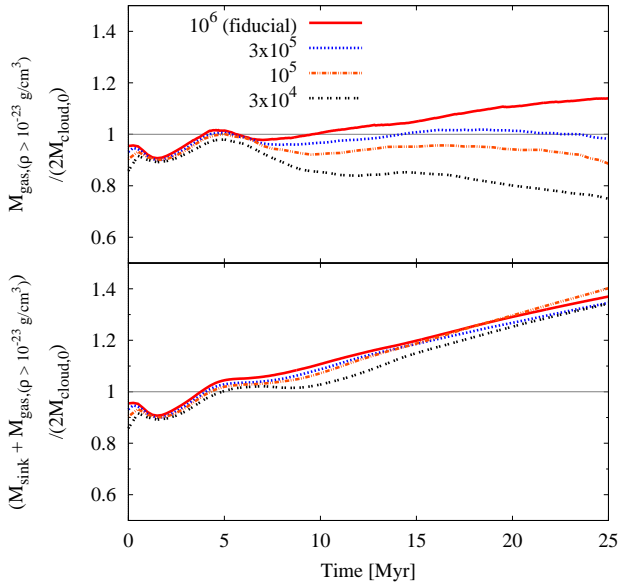}
\caption{Evolution of the dense gas mass (top) and combined dense gas mass and sink mass (bottom) as a fraction of the total initial cloud masses, as in \figref{fig:GasMassRatio}. Models with increasing resolution are better able to replenish the dense gas from the pristine background, while lower resolution models have sinks that are better able to accrete dense gas.}
\label{fig:GasMassRatio:res}
\end{figure}

\figref{fig:GasMassRatio:res} shows the total mass of the dense gas, with and without the sink particle mass.
The top panel shows that the amount of dense gas decreases for decreasing resolution as the simulations evolve.  This suggests that the background gas is less important at lower resolutions since it is unable to replenish the dense gas that is accreted onto the sinks.  

\begin{figure} 
\centering
\includegraphics[width=0.85\columnwidth]{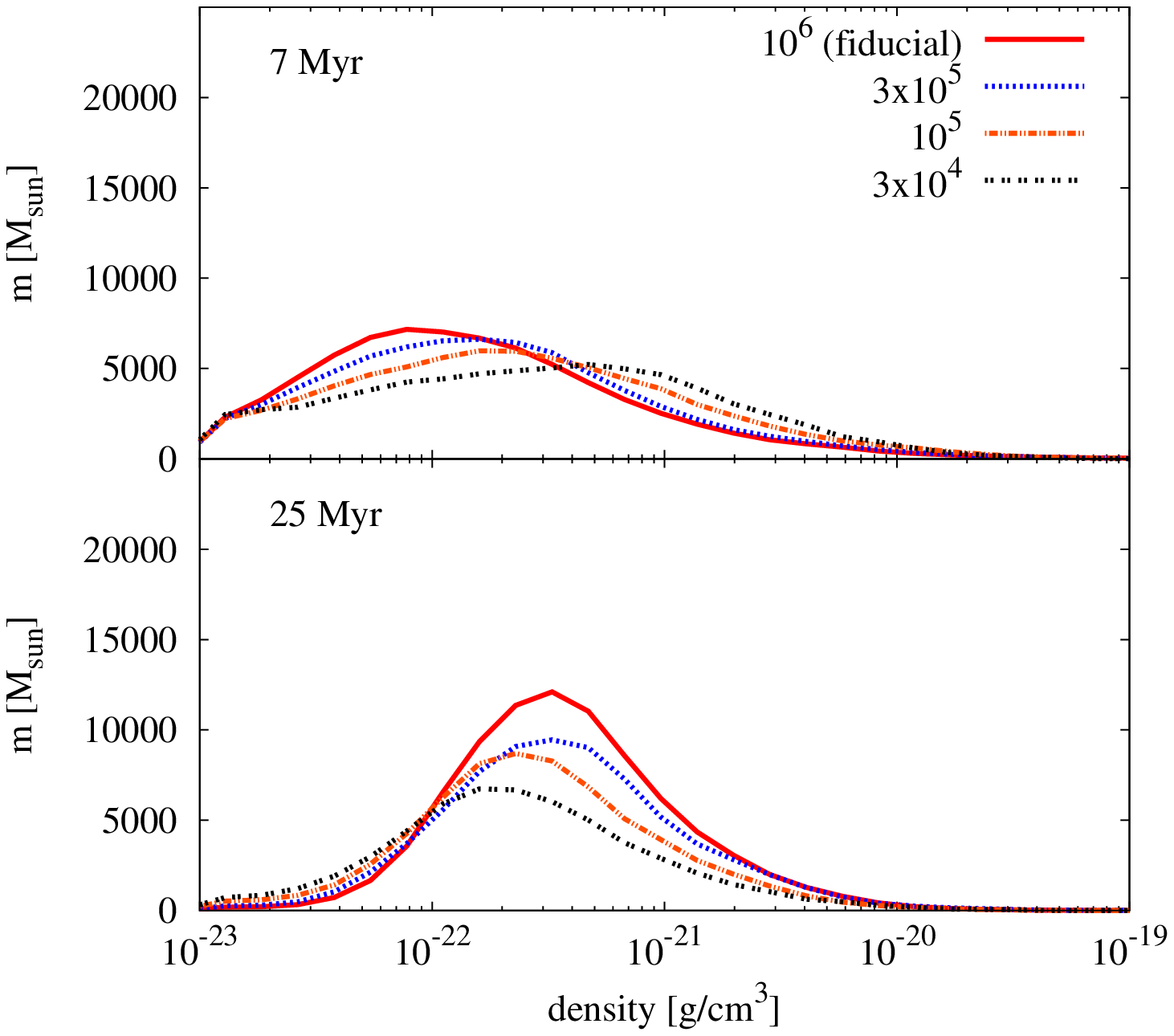}
\caption{Distribution of the gas densities at two times for \rhoge{-23} at four resolutions, as in \figref{fig:gasIMF}.  At 25~Myr, lower resolution models have less dense gas (and a lower peak density) since sink particles preferentially accrete high-density gas, depleting the high end of the distribution.}
\label{fig:gasIMF:res}
\end{figure}

In addition to the decreased amount of dense gas, the gas density distribution differs amongst resolutions, as shown in \figref{fig:gasIMF:res}.
At 7~Myr, all models have a similar total mass of dense gas (although \fidl{} is slightly lower).  The peak density decreases for increasing resolution, indicating that the initial collision has formed many more dense regions at lower resolutions.  As the system evolves, the relative distribution remains, but with more mass at lower densities for lower resolutions; these distributions are explicitly linked to star formation, as discussed below.  The clump mass function has a similar distribution, with the distribution shifting to higher masses for lower resolutions.

\begin{figure} 
\centering
\includegraphics[width=\columnwidth]{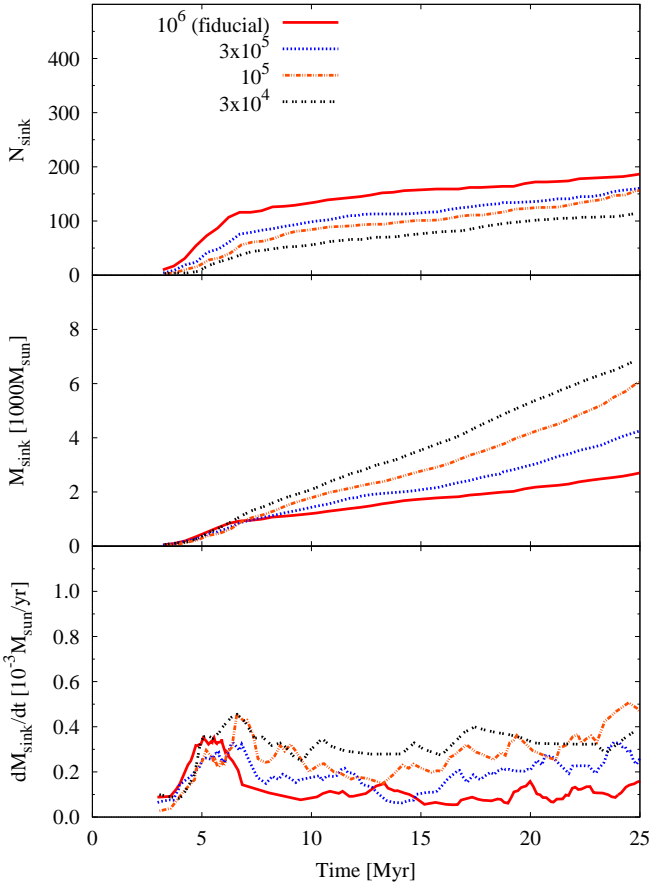}
\caption{From top to bottom: The total number of sink particles, the total mass in stars, and the total mass accretion rate onto sinks formation rate (moving averaged over 1~Myr), as in \figref{fig:sinkNME}. Although the sink accretion radius is the same for all resolutions, higher resolution models form more sinks, yet lower resolution models accrete more gas yielding more total mass in the sinks.  }
\label{fig:sinkNME:res}
\end{figure}

When considering the gas and sink mass (bottom panel of \figref{fig:GasMassRatio:res}), the sum is similar at all resolutions (differing at most by \sm8 per cent, occurring near \sm10~Myr), indicating that the sinks in the low resolution simulations are more efficient at accreting the dense gas.  This is better highlighted in \figref{fig:sinkNME:res} which shows the total number of sinks, total mass in sinks, and the sink accretion rate as a function of time.
At 25~Myr, there are nearly twice as many sinks in \fid{} than \fidl{}. Sinks are inserted when the particles within 2$h$ of the candidate particle meet given criteria \citep{BateBonnellPrice1995}.  The smoothing length, $h$, and hence the total mass within 2$h$ is resolution-dependent\footnote{In SPH simulations, there are \sm58 particles within 2$h$ when using the $M_4$ cubic spline kernel.}, and lower-resolution clumps tend to meet the given star-forming criteria and form sinks at lower maximum densities than in higher resolution simulations, in part due to the less well-defined velocities.  Therefore, the slightly earlier formation time might be able to affect the dynamics of the environment.  However, given that all sink particles in all simulations have the same accretion radius, it is unlikely that a slightly younger sink would be able to affect subsequent sink formation to the extent show in \figref{fig:sinkNME:res}.  Therefore, we conclude that the higher sink formation rate in \fid{} is a result of the gas behaviour, namely gas clumping.  In agreement with the literature, we conclude that higher resolution simulations are more prone to fragmentation than low resolutions simulations.  

\begin{figure} 
\centering
\includegraphics[width=\columnwidth]{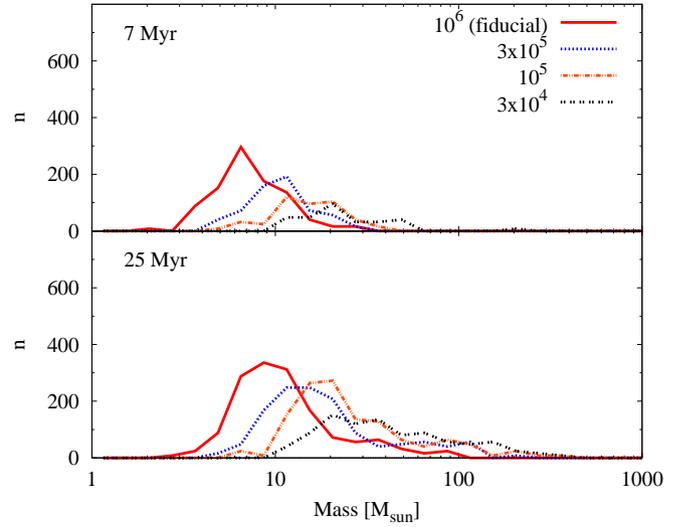}
\caption{Sink mass functions at two different epochs, as in \figref{fig:sinkSMF}; the bins are evenly spaced in log-space.  At both times, the mass function shifts to higher masses for decreasing resolution since sinks in lower resolution simulations more efficiently accrete gas.}
\label{fig:sinkSMF:res}
\end{figure}

Once the sinks have formed, the sinks in the lower resolution simulations accrete more gas, yielding higher total sink mass (second panel of \figref{fig:sinkNME:res}); this accretion is higher at all times for lower resolutions (bottom panel).  Although the sink accretion radius is the same at all resolutions, the particles at lower resolutions can more easily pass the required criteria to accrete onto a sink since their properties are less well-resolved; this also explains why there is a higher fraction of clumps that contain only one sink and no gas (20.0 and 10.7 per cent for \fidl{} and \fid{}, respectively).  Moreover, since there are fewer clumps forming, it is easier for the existing sinks to accrete more gas since there is less competition.  Indeed, the entire sink mass function is shifted to higher masses at lower resolutions; see \figref{fig:sinkSMF:res}.  
Additionally, sinks preferentially accrete higher density gas, thus the sinks in the lower resolution models better deplete the high-density end of the gas distribution function by 25~Myr (\figref{fig:gasIMF:res}) which accounts for the decreasing peak density for decreasing resolution.  

Based upon these results, colliding cloud / flow simulations are highly susceptible to resolution effects, especially if the conclusions are based upon the number and property of the sink particles.  Similar warnings about sink particles are discussed in (e.g.,) \citet{MachidaInutsukaMatsumoto2014} and \citet{Dobbs+2022}, where they show the impact of modifying sink parameters (although at a constant numerical resolution).  Therefore, caution must be applied to these results since we clearly have not reached numerical convergence; caution should be applied to similar results in the literature that have not likewise not demonstrated numerical convergence\footnote{Note that \citet{TakahiraTaskerHabe2014} disagree, since they obtained very similar results at both of their resolutions, which differed by a factor of two.}.

\section{Summary and conclusion}
\label{sec:conc}
In this paper, we model the collision of two elliptical clouds impacting at a default angle of 90$^\circ$.  The collision occurs in a warm background medium, where the initial cloud and background temperatures and densities were chosen to sit on the cooling curve of \citet{KoyamaInutsuka2002} to straddle the unstable equilibrium point at $n \approx1$~\percc{}; the initial densities differ by a factor of \appx10.  The clouds and the background have the same initial velocity in the $x$-direction, and the remnant retains this motion after the collision. We performed 11 simulations to determine how the initial properties (including cloud mass, velocity, internal turbulence, and impact angle) affected the remnant.  Our main results are as follows.

\begin{enumerate}
\item Our remnants are highly filamentary, but do not display the traditional hub-filament network; in general, gas flows outward from the collision point to form  filaments rather than along filaments to converge at hubs.  Clumps form from gravitationally collapsing over-densities within the filaments.
\item The primary property responsible for determining the size of the remnant is its initial velocity and its velocity-components.  The initial internal turbulence only plays an appreciable role when it is relatively large (as in our most turbulent model \mmachvhigh{}).
\item Despite forming stars, the total amount of dense gas increases (except in \mangvlow{}) as the remnant propagates through the background and accretes the pristine gas.  Therefore, the background is dynamically important to replenish the reservoir of dense gas.
\item After the collision, the velocity distribution of the gas narrows as the remnant evolves since the background gas hinders motion that is not equal to the background gas.  Although the initial $y$-velocity distribution is bi-modal, the collision reorders it into a uni-modal distribution with a decreasing width as the remnant's expansion in the $y$-direction slows.
\item The final number of sink particles (representing star clusters) varies by a factor of a few amongst the models by the end of the simulation.  The sink mass distributions are similar for all models, indicating that the initial conditions play a minimal role.  Most sinks represent star clusters, and each model contains at least a few sinks massive enough to contain one or more high-mass stars.  The outlier in the mass distribution is \mangvlow{}, which has fewer sinks than the remaining models, but these sinks are generally more massive; this is a result of the remnant being relatively confined to the $y \approx 0$ plane.
\item The distribution of properties (except velocity) of the resulting clumps (gas-only, sink-only, or gas+sink) is approximately independent of the initial conditions when considering normalised distributions. 
\item Colliding cloud simulations are highly resolution-dependent.  The number of star clusters (sink particles) increases with increasing resolution, but the mass of the star clusters decreases; the total stellar mass is approximately constant amongst the resolutions.  
\end{enumerate}

From the initial parameters investigated, the initial velocity and its components are the most important properties in determining the properties of the remnant.  However, in general, we have shown that when modelling hydrodynamic cloud collisions in a relatively dense background, the initial conditions play a minimal role in the formation of the remnant and its properties.

\section*{Acknowledgements}
We would like to thank the anonymous referee for useful comments that improved the quality of this manuscript.
We would like to thank Daniel J. Price and Steven Rieder for useful discussions regarding cooling.  
We would like to thank Katerina Klos for useful discussions regarding clump-finding.  
JW and IAB acknowledge support from the University of St Andrews.
This work was performed using the DiRAC Data Intensive service at Leicester, operated by the University of Leicester IT Services, which forms part of the STFC DiRAC HPC Facility (www.dirac.ac.uk). The equipment was funded by BEIS capital funding via STFC capital grants ST/K000373/1 and ST/R002363/1 and STFC DiRAC Operations grant ST/R001014/1. DiRAC is part of the National e-Infrastructure.
This works was also performed on the Kennedy computer cluster hosted at the University of St. Andrews.
In order to meet institutional and research funder open access requirements, any accepted manuscript arising shall be open access under a Creative Commons Attribution (CC BY) reuse licence with zero embargo.  
Several figures were made using \textsc{splash} \citep{Price2007}.  

\section*{Data availability}

The data underlying this article will be available upon reasonable request.  \textsc{Phantom} \citep{Price+2018phantom}, including the modifications designed for this study, is publicly available at \url{https://github.com/danieljprice/phantom}.  \textsc{Splash} \citep{Price2007} is publicly available at \url{https://github.com/danieljprice/splash}.

\bibliography{CloudCollision.bib}

\begin{thebibliography}{}
\makeatletter
\relax
\def\mn@urlcharsother{\let\do\@makeother \do\$\do\&\do\#\do\^\do\_\do\%\do\~}
\def\mn@doi{\begingroup\mn@urlcharsother \@ifnextchar [ {\mn@doi@}
  {\mn@doi@[]}}
\def\mn@doi@[#1]#2{\def\@tempa{#1}\ifx\@tempa\@empty \href
  {http://dx.doi.org/#2} {doi:#2}\else \href {http://dx.doi.org/#2} {#1}\fi
  \endgroup}
\def\mn@eprint#1#2{\mn@eprint@#1:#2::\@nil}
\def\mn@eprint@arXiv#1{\href {http://arxiv.org/abs/#1} {{\tt arXiv:#1}}}
\def\mn@eprint@dblp#1{\href {http://dblp.uni-trier.de/rec/bibtex/#1.xml}
  {dblp:#1}}
\def\mn@eprint@#1:#2:#3:#4\@nil{\def\@tempa {#1}\def\@tempb {#2}\def\@tempc
  {#3}\ifx \@tempc \@empty \let \@tempc \@tempb \let \@tempb \@tempa \fi \ifx
  \@tempb \@empty \def\@tempb {arXiv}\fi \@ifundefined
  {mn@eprint@\@tempb}{\@tempb:\@tempc}{\expandafter \expandafter \csname
  mn@eprint@\@tempb\endcsname \expandafter{\@tempc}}}

\bibitem[\protect\citeauthoryear{{Anderson} et~al.,}{{Anderson}
  et~al.}{2021}]{Anderson+2021}
{Anderson} M.,  et~al., 2021, \mn@doi [\mnras] {10.1093/mnras/stab2674}, \href
  {https://ui.adsabs.harvard.edu/abs/2021MNRAS.508.2964A} {508, 2964}

\bibitem[\protect\citeauthoryear{{Balfour}, {Whitworth}, {Hubber}  \&
  {Jaffa}}{{Balfour} et~al.}{2015}]{Balfour+2015}
{Balfour} S.~K.,  {Whitworth} A.~P.,  {Hubber} D.~A.,   {Jaffa} S.~E.,  2015,
  \mn@doi [\mnras] {10.1093/mnras/stv1772}, \href
  {https://ui.adsabs.harvard.edu/abs/2015MNRAS.453.2471B} {453, 2471}

\bibitem[\protect\citeauthoryear{{Balfour}, {Whitworth}  \& {Hubber}}{{Balfour}
  et~al.}{2017}]{BalfourWhitworthHubber2017}
{Balfour} S.~K.,  {Whitworth} A.~P.,   {Hubber} D.~A.,  2017, \mn@doi [\mnras]
  {10.1093/mnras/stw2956}, \href
  {https://ui.adsabs.harvard.edu/abs/2017MNRAS.465.3483B} {465, 3483}

\bibitem[\protect\citeauthoryear{{Bate}}{{Bate}}{2012}]{Bate2012}
{Bate} M.~R.,  2012, \mn@doi [\mnras] {10.1111/j.1365-2966.2011.19955.x}, \href
  {http://adsabs.harvard.edu/abs/2012MNRAS.419.3115B} {419, 3115}

\bibitem[\protect\citeauthoryear{{Bate}}{{Bate}}{2018}]{Bate2018}
{Bate} M.~R.,  2018, \mn@doi [\mnras] {10.1093/mnras/sty169}, \href
  {http://adsabs.harvard.edu/abs/2018MNRAS.475.5618B} {475, 5618}

\bibitem[\protect\citeauthoryear{{Bate}, {Bonnell}  \& {Price}}{{Bate}
  et~al.}{1995}]{BateBonnellPrice1995}
{Bate} M.~R.,  {Bonnell} I.~A.,   {Price} N.~M.,  1995, \mn@doi [\mnras]
  {10.1093/mnras/277.2.362}, \href
  {http://adsabs.harvard.edu/abs/1995MNRAS.277..362B} {277, 362}

\bibitem[\protect\citeauthoryear{{Bate}, {Bonnell}  \& {Bromm}}{{Bate}
  et~al.}{2003}]{BateBonnellBromm2003}
{Bate} M.~R.,  {Bonnell} I.~A.,   {Bromm} V.,  2003, \mn@doi [\mnras]
  {10.1046/j.1365-8711.2003.06210.x}, \href
  {http://adsabs.harvard.edu/abs/2003MNRAS.339..577B} {339, 577}

\bibitem[\protect\citeauthoryear{{Beltr{\'a}n}, {Rivilla}, {Kumar}, {Cesaroni}
  \& {Galli}}{{Beltr{\'a}n} et~al.}{2022}]{Beltran+2022}
{Beltr{\'a}n} M.~T.,  {Rivilla} V.~M.,  {Kumar} M.~S.~N.,  {Cesaroni} R.,
  {Galli} D.,  2022, \mn@doi [\aap] {10.1051/0004-6361/202243361}, \href
  {https://ui.adsabs.harvard.edu/abs/2022A&A...660L...4B} {660, L4}

\bibitem[\protect\citeauthoryear{{Bonnell}, {Bate}, {Clarke}  \&
  {Pringle}}{{Bonnell} et~al.}{1997}]{Bonnell+1997}
{Bonnell} I.~A.,  {Bate} M.~R.,  {Clarke} C.~J.,   {Pringle} J.~E.,  1997,
  \mn@doi [\mnras] {10.1093/mnras/285.1.201}, \href
  {https://ui.adsabs.harvard.edu/abs/1997MNRAS.285..201B} {285, 201}

\bibitem[\protect\citeauthoryear{{Bonnell}, {Bate}, {Clarke}  \&
  {Pringle}}{{Bonnell} et~al.}{2001}]{Bonnell+2001ca}
{Bonnell} I.~A.,  {Bate} M.~R.,  {Clarke} C.~J.,   {Pringle} J.~E.,  2001,
  \mn@doi [\mnras] {10.1046/j.1365-8711.2001.04270.x}, \href
  {https://ui.adsabs.harvard.edu/abs/2001MNRAS.323..785B} {323, 785}

\bibitem[\protect\citeauthoryear{{Carroll-Nellenback}, {Frank}  \&
  {Heitsch}}{{Carroll-Nellenback}
  et~al.}{2014}]{CarrollnellenbackFrankHeitsch2014}
{Carroll-Nellenback} J.~J.,  {Frank} A.,   {Heitsch} F.,  2014, \mn@doi [\apj]
  {10.1088/0004-637X/790/1/37}, \href
  {https://ui.adsabs.harvard.edu/abs/2014ApJ...790...37C} {790, 37}

\bibitem[\protect\citeauthoryear{{Clark}, {Glover}, {Ragan}  \&
  {Duarte-Cabral}}{{Clark} et~al.}{2019}]{Clark+2019}
{Clark} P.~C.,  {Glover} S. C.~O.,  {Ragan} S.~E.,   {Duarte-Cabral} A.,  2019,
  \mn@doi [\mnras] {10.1093/mnras/stz1119}, \href
  {https://ui.adsabs.harvard.edu/abs/2019MNRAS.486.4622C} {486, 4622}

\bibitem[\protect\citeauthoryear{{Colombo} et~al.,}{{Colombo}
  et~al.}{2014}]{Colombo+2014}
{Colombo} D.,  et~al., 2014, \mn@doi [\apj] {10.1088/0004-637X/784/1/3}, \href
  {https://ui.adsabs.harvard.edu/abs/2014ApJ...784....3C} {784, 3}

\bibitem[\protect\citeauthoryear{{Crutcher}}{{Crutcher}}{2012}]{Crutcher2012}
{Crutcher} R.~M.,  2012, \mn@doi [\araa] {10.1146/annurev-astro-081811-125514},
  \href {http://adsabs.harvard.edu/abs/2012ARA&A..50...29C} {50, 29}

\bibitem[\protect\citeauthoryear{{Dobbs} \& {Wurster}}{{Dobbs} \&
  {Wurster}}{2021}]{DobbsWurster2021}
{Dobbs} C.~L.,  {Wurster} J.,  2021, \mn@doi [\mnras] {10.1093/mnras/stab150},
  \href {https://ui.adsabs.harvard.edu/abs/2021MNRAS.502.2285D} {502, 2285}

\bibitem[\protect\citeauthoryear{{Dobbs}, {Pringle}  \&
  {Duarte-Cabral}}{{Dobbs} et~al.}{2015}]{DobbsPringleDuartecabral2015}
{Dobbs} C.~L.,  {Pringle} J.~E.,   {Duarte-Cabral} A.,  2015, \mn@doi [\mnras]
  {10.1093/mnras/stu2319}, \href
  {https://ui.adsabs.harvard.edu/abs/2015MNRAS.446.3608D} {446, 3608}

\bibitem[\protect\citeauthoryear{{Dobbs}, {Bending}, {Pettitt}  \&
  {Bate}}{{Dobbs} et~al.}{2022}]{Dobbs+2022}
{Dobbs} C.~L.,  {Bending} T.~J.~R.,  {Pettitt} A.~R.,   {Bate} M.~R.,  2022,
  \mn@doi [\mnras] {10.1093/mnras/stab3036}, \href
  {https://ui.adsabs.harvard.edu/abs/2022MNRAS.509..954D} {509, 954}

\bibitem[\protect\citeauthoryear{{Duarte-Cabral} \& {Dobbs}}{{Duarte-Cabral} \&
  {Dobbs}}{2016}]{DuartecabralDobbs2016}
{Duarte-Cabral} A.,  {Dobbs} C.~L.,  2016, \mn@doi [\mnras]
  {10.1093/mnras/stw469}, \href
  {https://ui.adsabs.harvard.edu/abs/2016MNRAS.458.3667D} {458, 3667}

\bibitem[\protect\citeauthoryear{{Federrath} \& {Klessen}}{{Federrath} \&
  {Klessen}}{2012}]{FederrathKlessen2012}
{Federrath} C.,  {Klessen} R.~S.,  2012, \mn@doi [\apj]
  {10.1088/0004-637X/761/2/156}, \href
  {https://ui.adsabs.harvard.edu/abs/2012ApJ...761..156F} {761, 156}

\bibitem[\protect\citeauthoryear{{Federrath} \& {Klessen}}{{Federrath} \&
  {Klessen}}{2013}]{FederrathKlessen2013}
{Federrath} C.,  {Klessen} R.~S.,  2013, \mn@doi [\apj]
  {10.1088/0004-637X/763/1/51}, \href
  {http://adsabs.harvard.edu/abs/2013ApJ...763...51F} {763, 51}

\bibitem[\protect\citeauthoryear{{Fogerty}, {Frank}, {Heitsch},
  {Carroll-Nellenback}, {Haig}  \& {Adams}}{{Fogerty}
  et~al.}{2016}]{Fogerty+2016}
{Fogerty} E.,  {Frank} A.,  {Heitsch} F.,  {Carroll-Nellenback} J.,  {Haig} C.,
    {Adams} M.,  2016, \mn@doi [\mnras] {10.1093/mnras/stw1141}, \href
  {https://ui.adsabs.harvard.edu/abs/2016MNRAS.460.2110F} {460, 2110}

\bibitem[\protect\citeauthoryear{{Fujita} et~al.,}{{Fujita}
  et~al.}{2021}]{Fujita+2021}
{Fujita} S.,  et~al., 2021, \mn@doi [\pasj] {10.1093/pasj/psaa005}, \href
  {https://ui.adsabs.harvard.edu/abs/2021PASJ...73S.273F} {73, S273}

\bibitem[\protect\citeauthoryear{{Fukui} et~al.,}{{Fukui}
  et~al.}{2014}]{Fukui+2014}
{Fukui} Y.,  et~al., 2014, \mn@doi [\apj] {10.1088/0004-637X/780/1/36}, \href
  {https://ui.adsabs.harvard.edu/abs/2014ApJ...780...36F} {780, 36}

\bibitem[\protect\citeauthoryear{{Fukui} et~al.,}{{Fukui}
  et~al.}{2016}]{Fukui+2016}
{Fukui} Y.,  et~al., 2016, \mn@doi [\apj] {10.3847/0004-637X/820/1/26}, \href
  {https://ui.adsabs.harvard.edu/abs/2016ApJ...820...26F} {820, 26}

\bibitem[\protect\citeauthoryear{{Fukui}, {Tsuge}, {Sano}, {Bekki}, {Yozin},
  {Tachihara}  \& {Inoue}}{{Fukui} et~al.}{2017}]{Fukui+2017}
{Fukui} Y.,  {Tsuge} K.,  {Sano} H.,  {Bekki} K.,  {Yozin} C.,  {Tachihara} K.,
    {Inoue} T.,  2017, \mn@doi [\pasj] {10.1093/pasj/psx032}, \href
  {https://ui.adsabs.harvard.edu/abs/2017PASJ...69L...5F} {69, L5}

\bibitem[\protect\citeauthoryear{{Fukui} et~al.,}{{Fukui}
  et~al.}{2018a}]{Fukui+2018gm}
{Fukui} Y.,  et~al., 2018a, \mn@doi [\pasj] {10.1093/pasj/psx144}, \href
  {https://ui.adsabs.harvard.edu/abs/2018PASJ...70S..44F} {70, S44}

\bibitem[\protect\citeauthoryear{{Fukui} et~al.,}{{Fukui}
  et~al.}{2018b}]{Fukui+2018orion}
{Fukui} Y.,  et~al., 2018b, \mn@doi [\apj] {10.3847/1538-4357/aac217}, \href
  {https://ui.adsabs.harvard.edu/abs/2018ApJ...859..166F} {859, 166}

\bibitem[\protect\citeauthoryear{{Fukui}, {Habe}, {Inoue}, {Enokiya}  \&
  {Tachihara}}{{Fukui} et~al.}{2021}]{Fukui+2021}
{Fukui} Y.,  {Habe} A.,  {Inoue} T.,  {Enokiya} R.,   {Tachihara} K.,  2021,
  \mn@doi [\pasj] {10.1093/pasj/psaa103}, \href
  {https://ui.adsabs.harvard.edu/abs/2021PASJ...73S...1F} {73, S1}

\bibitem[\protect\citeauthoryear{{Furukawa}, {Dawson}, {Ohama}, {Kawamura},
  {Mizuno}, {Onishi}  \& {Fukui}}{{Furukawa} et~al.}{2009}]{Furukawa+2009}
{Furukawa} N.,  {Dawson} J.~R.,  {Ohama} A.,  {Kawamura} A.,  {Mizuno} N.,
  {Onishi} T.,   {Fukui} Y.,  2009, \mn@doi [\apjl]
  {10.1088/0004-637X/696/2/L115}, \href
  {https://ui.adsabs.harvard.edu/abs/2009ApJ...696L.115F} {696, L115}

\bibitem[\protect\citeauthoryear{{Glover} \& {Mac Low}}{{Glover} \& {Mac
  Low}}{2007}]{GloverMaclow2007}
{Glover} S. C.~O.,  {Mac Low} M.-M.,  2007, \mn@doi [\apjs] {10.1086/512238},
  \href {https://ui.adsabs.harvard.edu/abs/2007ApJS..169..239G} {169, 239}

\bibitem[\protect\citeauthoryear{{Haworth} et~al.,}{{Haworth}
  et~al.}{2015a}]{Haworth+2015bridgeA}
{Haworth} T.~J.,  et~al., 2015a, \mn@doi [\mnras] {10.1093/mnras/stv639}, \href
  {https://ui.adsabs.harvard.edu/abs/2015MNRAS.450...10H} {450, 10}

\bibitem[\protect\citeauthoryear{{Haworth}, {Shima}, {Tasker}, {Fukui},
  {Torii}, {Dale}, {Takahira}  \& {Habe}}{{Haworth}
  et~al.}{2015b}]{Haworth+2015bridgeB}
{Haworth} T.~J.,  {Shima} K.,  {Tasker} E.~J.,  {Fukui} Y.,  {Torii} K.,
  {Dale} J.~E.,  {Takahira} K.,   {Habe} A.,  2015b, \mn@doi [\mnras]
  {10.1093/mnras/stv2068}, \href
  {https://ui.adsabs.harvard.edu/abs/2015MNRAS.454.1634H} {454, 1634}

\bibitem[\protect\citeauthoryear{{Heyer} \& {Brunt}}{{Heyer} \&
  {Brunt}}{2004}]{HeyerBrunt2004}
{Heyer} M.~H.,  {Brunt} C.~M.,  2004, \mn@doi [\apjl] {10.1086/425978}, \href
  {http://adsabs.harvard.edu/abs/2004ApJ...615L..45H} {615, L45}

\bibitem[\protect\citeauthoryear{{Higuchi}, {Chibueze}, {Habe}, {Takahira}  \&
  {Takano}}{{Higuchi} et~al.}{2014}]{Higuchi+2014}
{Higuchi} A.~E.,  {Chibueze} J.~O.,  {Habe} A.,  {Takahira} K.,   {Takano} S.,
  2014, \mn@doi [\aj] {10.1088/0004-6256/147/6/141}, \href
  {https://ui.adsabs.harvard.edu/abs/2014AJ....147..141H} {147, 141}

\bibitem[\protect\citeauthoryear{{Inoue} \& {Fukui}}{{Inoue} \&
  {Fukui}}{2013}]{InoueFukui2013}
{Inoue} T.,  {Fukui} Y.,  2013, \mn@doi [\apjl] {10.1088/2041-8205/774/2/L31},
  \href {https://ui.adsabs.harvard.edu/abs/2013ApJ...774L..31I} {774, L31}

\bibitem[\protect\citeauthoryear{{Kinoshita} \& {Nakamura}}{{Kinoshita} \&
  {Nakamura}}{2022}]{KinoshitaNakamura2022}
{Kinoshita} S.~W.,  {Nakamura} F.,  2022, \mn@doi [\apj]
  {10.3847/1538-4357/ac8c95}, \href
  {https://ui.adsabs.harvard.edu/abs/2022ApJ...937...69K} {937, 69}

\bibitem[\protect\citeauthoryear{{Koyama} \& {Inutsuka}}{{Koyama} \&
  {Inutsuka}}{2002}]{KoyamaInutsuka2002}
{Koyama} H.,  {Inutsuka} S.-i.,  2002, \mn@doi [\apjl] {10.1086/338978}, \href
  {https://ui.adsabs.harvard.edu/abs/2002ApJ...564L..97K} {564, L97}

\bibitem[\protect\citeauthoryear{{Kumar}, {Palmeirim}, {Arzoumanian}  \&
  {Inutsuka}}{{Kumar} et~al.}{2020}]{Kumar+2020}
{Kumar} M.~S.~N.,  {Palmeirim} P.,  {Arzoumanian} D.,   {Inutsuka} S.~I.,
  2020, \mn@doi [\aap] {10.1051/0004-6361/202038232}, \href
  {https://ui.adsabs.harvard.edu/abs/2020A&A...642A..87K} {642, A87}

\bibitem[\protect\citeauthoryear{{Liow} \& {Dobbs}}{{Liow} \&
  {Dobbs}}{2020}]{LiowDobbs2020}
{Liow} K.~Y.,  {Dobbs} C.~L.,  2020, \mn@doi [\mnras] {10.1093/mnras/staa2857},
  \href {https://ui.adsabs.harvard.edu/abs/2020MNRAS.499.1099L} {499, 1099}

\bibitem[\protect\citeauthoryear{{Machida}, {Inutsuka}  \&
  {Matsumoto}}{{Machida} et~al.}{2014}]{MachidaInutsukaMatsumoto2014}
{Machida} M.~N.,  {Inutsuka} S.-i.,   {Matsumoto} T.,  2014, \mn@doi [\mnras]
  {10.1093/mnras/stt2343}, \href
  {http://adsabs.harvard.edu/abs/2014MNRAS.438.2278M} {438, 2278}

\bibitem[\protect\citeauthoryear{{Matsui} et~al.,}{{Matsui}
  et~al.}{2012}]{Matsui+2012}
{Matsui} H.,  et~al., 2012, \mn@doi [\apj] {10.1088/0004-637X/746/1/26}, \href
  {https://ui.adsabs.harvard.edu/abs/2012ApJ...746...26M} {746, 26}

\bibitem[\protect\citeauthoryear{{Matsui}, {Tanikawa}  \& {Saitoh}}{{Matsui}
  et~al.}{2019}]{MatsuiTanikawaSaitoh2019}
{Matsui} H.,  {Tanikawa} A.,   {Saitoh} T.~R.,  2019, \mn@doi [\pasj]
  {10.1093/pasj/psy139}, \href
  {https://ui.adsabs.harvard.edu/abs/2019PASJ...71...19M} {71, 19}

\bibitem[\protect\citeauthoryear{{McKee} \& {Tan}}{{McKee} \&
  {Tan}}{2003}]{MckeeTan2003}
{McKee} C.~F.,  {Tan} J.~C.,  2003, \mn@doi [\apj] {10.1086/346149}, \href
  {https://ui.adsabs.harvard.edu/abs/2003ApJ...585..850M} {585, 850}

\bibitem[\protect\citeauthoryear{{Meru} \& {Bate}}{{Meru} \&
  {Bate}}{2012}]{MeruBate2012}
{Meru} F.,  {Bate} M.~R.,  2012, \mn@doi [\mnras]
  {10.1111/j.1365-2966.2012.22035.x}, \href
  {http://adsabs.harvard.edu/abs/2012MNRAS.427.2022M} {427, 2022}

\bibitem[\protect\citeauthoryear{{Meyer}, {Kuiper}, {Kley}, {Johnston}  \&
  {Vorobyov}}{{Meyer} et~al.}{2018}]{Meyer+2018}
{Meyer} D.~M.-A.,  {Kuiper} R.,  {Kley} W.,  {Johnston} K.~G.,   {Vorobyov} E.,
   2018, \mn@doi [\mnras] {10.1093/mnras/stx2551}, \href
  {http://adsabs.harvard.edu/abs/2018MNRAS.473.3615M} {473, 3615}

\bibitem[\protect\citeauthoryear{{Micic}, {Glover}, {Banerjee}  \&
  {Klessen}}{{Micic} et~al.}{2013}]{Micic+2013}
{Micic} M.,  {Glover} S. C.~O.,  {Banerjee} R.,   {Klessen} R.~S.,  2013,
  \mn@doi [\mnras] {10.1093/mnras/stt489}, \href
  {https://ui.adsabs.harvard.edu/abs/2013MNRAS.432..626M} {432, 626}

\bibitem[\protect\citeauthoryear{{Myers}}{{Myers}}{2009}]{Myers2009}
{Myers} P.~C.,  2009, \mn@doi [\apj] {10.1088/0004-637X/700/2/1609}, \href
  {https://ui.adsabs.harvard.edu/abs/2009ApJ...700.1609M} {700, 1609}

\bibitem[\protect\citeauthoryear{{Ohama} et~al.,}{{Ohama}
  et~al.}{2010}]{Ohama+2010}
{Ohama} A.,  et~al., 2010, \mn@doi [\apj] {10.1088/0004-637X/709/2/975}, \href
  {https://ui.adsabs.harvard.edu/abs/2010ApJ...709..975O} {709, 975}

\bibitem[\protect\citeauthoryear{{Ostriker}, {Stone}  \& {Gammie}}{{Ostriker}
  et~al.}{2001}]{OstrikerStoneGammie2001}
{Ostriker} E.~C.,  {Stone} J.~M.,   {Gammie} C.~F.,  2001, \mn@doi [\apj]
  {10.1086/318290}, \href {http://adsabs.harvard.edu/abs/2001ApJ...546..980O}
  {546, 980}

\bibitem[\protect\citeauthoryear{{Padoan}, {Pan}, {Juvela}, {Haugb{\o}lle}  \&
  {Nordlund}}{{Padoan} et~al.}{2020}]{Padoan+2020}
{Padoan} P.,  {Pan} L.,  {Juvela} M.,  {Haugb{\o}lle} T.,   {Nordlund}
  {\r{A}}.,  2020, \mn@doi [\apj] {10.3847/1538-4357/abaa47}, \href
  {https://ui.adsabs.harvard.edu/abs/2020ApJ...900...82P} {900, 82}

\bibitem[\protect\citeauthoryear{{Peretto} et~al.,}{{Peretto}
  et~al.}{2013}]{Peretto+2013}
{Peretto} N.,  et~al., 2013, \mn@doi [\aap] {10.1051/0004-6361/201321318},
  \href {https://ui.adsabs.harvard.edu/abs/2013A&A...555A.112P} {555, A112}

\bibitem[\protect\citeauthoryear{{Price}}{{Price}}{2007}]{Price2007}
{Price} D.~J.,  2007, \mn@doi [\pasa] {10.1071/AS07022}, \href
  {http://adsabs.harvard.edu/abs/2007PASA...24..159P} {24, 159}

\bibitem[\protect\citeauthoryear{{Price} et~al.,}{{Price}
  et~al.}{2018}]{Price+2018phantom}
{Price} D.~J.,  et~al., 2018, \mn@doi [\pasa] {10.1017/pasa.2018.25}, \href
  {http://adsabs.harvard.edu/abs/2018PASA...35...31P} {35, e031}

\bibitem[\protect\citeauthoryear{{Priestley} \& {Whitworth}}{{Priestley} \&
  {Whitworth}}{2021}]{PriestleyWhitworth2021}
{Priestley} F.~D.,  {Whitworth} A.~P.,  2021, \mn@doi [\mnras]
  {10.1093/mnras/stab1777}, \href
  {https://ui.adsabs.harvard.edu/abs/2021MNRAS.506..775P} {506, 775}

\bibitem[\protect\citeauthoryear{Rocha, Velho  \& Carvalho}{Rocha
  et~al.}{2002}]{RochaVelhoCarvalho2002}
Rocha L.,  Velho L.,   Carvalho P. C.~P.,  2002, in 15th Brazilian Symposium on
  Computer Graphics and Image Processing {(SIBGRAPI} 2002), 7-10 October 2002,
  Fortaleza-CE, Brazil. {IEEE} Computer Society, pp 99--105,
  \mn@doi{10.1109/SIBGRA.2002.1167130}, \url
  {https://doi.org/10.1109/SIBGRA.2002.1167130}

\bibitem[\protect\citeauthoryear{{Sakre}, {Habe}, {Pettitt}  \&
  {Okamoto}}{{Sakre} et~al.}{2021}]{Sakre+2021}
{Sakre} N.,  {Habe} A.,  {Pettitt} A.~R.,   {Okamoto} T.,  2021, \mn@doi
  [\pasj] {10.1093/pasj/psaa059}, \href
  {https://ui.adsabs.harvard.edu/abs/2021PASJ...73S.385S} {73, S385}

\bibitem[\protect\citeauthoryear{{Sano} et~al.,}{{Sano}
  et~al.}{2021}]{Sano+2021}
{Sano} H.,  et~al., 2021, \mn@doi [\pasj] {10.1093/pasj/psaa045}, \href
  {https://ui.adsabs.harvard.edu/abs/2021PASJ...73S..62S} {73, S62}

\bibitem[\protect\citeauthoryear{{Smilgys} \& {Bonnell}}{{Smilgys} \&
  {Bonnell}}{2016}]{SmilgysBonnell2016}
{Smilgys} R.,  {Bonnell} I.~A.,  2016, \mn@doi [\mnras] {10.1093/mnras/stw791},
  \href {https://ui.adsabs.harvard.edu/abs/2016MNRAS.459.1985S} {459, 1985}

\bibitem[\protect\citeauthoryear{{Takahira}, {Tasker}  \& {Habe}}{{Takahira}
  et~al.}{2014}]{TakahiraTaskerHabe2014}
{Takahira} K.,  {Tasker} E.~J.,   {Habe} A.,  2014, \mn@doi [\apj]
  {10.1088/0004-637X/792/1/63}, \href
  {https://ui.adsabs.harvard.edu/abs/2014ApJ...792...63T} {792, 63}

\bibitem[\protect\citeauthoryear{{Tanvir} \& {Dale}}{{Tanvir} \&
  {Dale}}{2020}]{TanvirDale2020}
{Tanvir} T.~S.,  {Dale} J.~E.,  2020, \mn@doi [\mnras] {10.1093/mnras/staa665},
  \href {https://ui.adsabs.harvard.edu/abs/2020MNRAS.494..246T} {494, 246}

\bibitem[\protect\citeauthoryear{{Tanvir} \& {Dale}}{{Tanvir} \&
  {Dale}}{2021}]{TanvirDale2021}
{Tanvir} T.~S.,  {Dale} J.~E.,  2021, \mn@doi [\mnras]
  {10.1093/mnras/stab1389}, \href
  {https://ui.adsabs.harvard.edu/abs/2021MNRAS.506..824T} {506, 824}

\bibitem[\protect\citeauthoryear{{Tasker}}{{Tasker}}{2011}]{Tasker2011}
{Tasker} E.~J.,  2011, \mn@doi [\apj] {10.1088/0004-637X/730/1/11}, \href
  {https://ui.adsabs.harvard.edu/abs/2011ApJ...730...11T} {730, 11}

\bibitem[\protect\citeauthoryear{{Tasker} \& {Tan}}{{Tasker} \&
  {Tan}}{2009}]{TaskerTan2009}
{Tasker} E.~J.,  {Tan} J.~C.,  2009, \mn@doi [\apj]
  {10.1088/0004-637X/700/1/358}, \href
  {https://ui.adsabs.harvard.edu/abs/2009ApJ...700..358T} {700, 358}

\bibitem[\protect\citeauthoryear{{Tricco}, {Price}  \& {Federrath}}{{Tricco}
  et~al.}{2016}]{TriccoPriceFederrath2016}
{Tricco} T.~S.,  {Price} D.~J.,   {Federrath} C.,  2016, \mn@doi [\mnras]
  {10.1093/mnras/stw1280}, \href
  {http://adsabs.harvard.edu/abs/2016MNRAS.461.1260T} {461, 1260}

\bibitem[\protect\citeauthoryear{{V{\'a}zquez-Semadeni}, {G{\'o}mez},
  {Jappsen}, {Ballesteros-Paredes}, {Gonz{\'a}lez}  \&
  {Klessen}}{{V{\'a}zquez-Semadeni} et~al.}{2007}]{Vazquezsemadeni+2007}
{V{\'a}zquez-Semadeni} E.,  {G{\'o}mez} G.~C.,  {Jappsen} A.~K.,
  {Ballesteros-Paredes} J.,  {Gonz{\'a}lez} R.~F.,   {Klessen} R.~S.,  2007,
  \mn@doi [\apj] {10.1086/510771}, \href
  {https://ui.adsabs.harvard.edu/abs/2007ApJ...657..870V} {657, 870}

\bibitem[\protect\citeauthoryear{{V{\'a}zquez-Semadeni}, {Banerjee},
  {G{\'o}mez}, {Hennebelle}, {Duffin}  \& {Klessen}}{{V{\'a}zquez-Semadeni}
  et~al.}{2011}]{Vazquezsemadeni+2011}
{V{\'a}zquez-Semadeni} E.,  {Banerjee} R.,  {G{\'o}mez} G.~C.,  {Hennebelle}
  P.,  {Duffin} D.,   {Klessen} R.~S.,  2011, \mn@doi [\mnras]
  {10.1111/j.1365-2966.2011.18569.x}, \href
  {https://ui.adsabs.harvard.edu/abs/2011MNRAS.414.2511V} {414, 2511}

\bibitem[\protect\citeauthoryear{Verlet}{Verlet}{1967}]{Verlet1967}
Verlet L.,  1967, \mn@doi [Phys. Rev.] {10.1103/PhysRev.159.98}, 159, 98

\bibitem[\protect\citeauthoryear{{Wurster}, {Bate}  \& {Price}}{{Wurster}
  et~al.}{2019}]{WursterBatePrice2019}
{Wurster} J.,  {Bate} M.~R.,   {Price} D.~J.,  2019, \mn@doi [\mnras]
  {10.1093/mnras/stz2215}, \href
  {https://ui.adsabs.harvard.edu/abs/2019MNRAS.489.1719W} {489, 1719}

\bibitem[\protect\citeauthoryear{{Yamada} et~al.,}{{Yamada}
  et~al.}{2021}]{Yamada+2021}
{Yamada} R.~I.,  et~al., 2021, \mn@doi [\pasj] {10.1093/pasj/psab050}, \href
  {https://ui.adsabs.harvard.edu/abs/2021PASJ...73..880Y} {73, 880}

\bibitem[\protect\citeauthoryear{{Zinnecker} \& {Yorke}}{{Zinnecker} \&
  {Yorke}}{2007}]{ZinneckerYorke2007}
{Zinnecker} H.,  {Yorke} H.~W.,  2007, \mn@doi [\araa]
  {10.1146/annurev.astro.44.051905.092549}, \href
  {https://ui.adsabs.harvard.edu/abs/2007ARA&A..45..481Z} {45, 481}

\bibitem[\protect\citeauthoryear{{Zucker}, {Battersby}  \& {Goodman}}{{Zucker}
  et~al.}{2018}]{ZuckerBattersbyGoodman2018}
{Zucker} C.,  {Battersby} C.,   {Goodman} A.,  2018, \mn@doi [\apj]
  {10.3847/1538-4357/aacc66}, \href
  {https://ui.adsabs.harvard.edu/abs/2018ApJ...864..153Z} {864, 153}

\makeatother
\end{thebibliography}
\appendix
\section{Dynamic boundaries}
\label{app:bdy}

\subsection{The algorithm}

To prevent excessive computational expense on the ambient background medium, we have developed dynamic, periodic boundaries.  These boundaries expand and contract based upon the dynamics of the `interesting' regions to always ensure that there is enough medium into which the interesting regions can expand, but not so much as to be computationally prohibitive.  Although this paper presents purely hydrodynamic models, dynamic boundaries were developed to be compatible with magnetohydrodynamic (MHD) simulations, thus we present the full MHD algorithm and test both low-resolution hydrodynamic and MHD simulations.

We define an initial domain around the region of interest.  The boundaries of this domain are periodic and given the same initial velocity as the background medium to minimise the number of corrections to the boundaries themselves.  The boundaries of the domain are inspected at every \dtmax{} to determine if the boundaries need adjusting, where \dtmax{} is the time between outputs.

Both density and velocity are used to determine the required location of the boundaries.  First, all particles with $\rho_i > \rho_\text{bdy}$ are tagged, where $\rho_\text{bdy}$ is a density threshold between the background and initial cloud (`interesting') densities; we set $\rho_\text{bdy} = 6.63\times10^{-24}$~\gpercc{} $\approx1.7$~\percc{}.  Next, particles further than $4h$ from the boundary are tagged if their velocity meets the following criteria,
\begin{equation}
\label{eq:vcriteria}
\text{is tagged} = \left\{ \begin{array}{l l} \left| 1 - \frac{v_{\text{bdy},a}}{v} \right | > 0.05  		    &  \text{if } v_{\text{bdy},a} \neq 0, \\ \\
                                                                \left|      \frac{v_a}{v} \right | > 0.05  		                     &  \text{if } v_{\text{bdy},a} =  0,
\end{array}\right.
\end{equation}
where $a \in \left\{x,y,z\right\}$, $v_{\text{bdy},a}$ is the velocity of the boundary in the $a$-th direction, and $v$ and $v_a$ are the particle's total velocity and velocity in the $a$-th direction, respectively. 

We then determine the required location of the boundary by finding the extreme values of all tagged particles via
\begin{subequations}
\label{eq:newbdy}
\begin{flalign}
a_\text{min}  &= \min\left[ a_\text{min}, a + N_\text{dt}\text{d}t_\text{max}\left(v_a  - v_{\text{bdy},a} - v_{\text{f},a} \right), a - d_{\text{bkg},a} \right], &\\
a_\text{max} &= \max\left[ a_\text{max}, a + N_\text{dt}\text{d}t_\text{max}\left(v_a  - v_{\text{bdy},a} + v_{\text{f},a}  \right),  a + d_{\text{bkg},a}\right]. &
\end{flalign}
\end{subequations}
For purely hydrodynamic simulations, $v_{\text{f},a} = c_\text{s}$, where $c_\text{s}$ is the sound speed\footnote{For purely hydrodynamic simulations, $v_{\text{f},a} \equiv 0$ is also valid.}.  In MHD simulations, fast magnetosonic waves rapidly travel away from the region of interest.  If left unchecked, these waves reach the boundaries, interact with themselves (since the boundaries are periodic) and cause artificial density and magnetic field enhancements.  Therefore, for MHD simulations, $v_{\text{f},a}$ is the fast magnetosonic wave given by 
\begin{equation}
v_{\text{f},a} = \sqrt{ \frac{1}{2} \left( v_\text{A}^2 + c_\text{s}^2 + \sqrt{\left(v_\text{A}^2 + c_\text{s}^2\right)^2 - 4\frac{B_a^2}{\rho} c_\text{s}^2} \right)},
\end{equation}
where $v_\text{A}$ is the \alfven{} velocity and $B_a$ is the magnetic field strength of the $a$-th component.  We set $N_\text{dt} = 3$; smaller values may lead to numerical artefacts if the wave speed increases rapidly between boundary updates.  In Eqn.~\ref{eq:newbdy}, we also set a minimum distance between the most extreme particle and the border, $d_{\text{bkg},a}$.  In the simulations presented in this paper, we liberally set $d_{\text{bkg},+x} = 200$~pc and $d_{\text{bkg},a} = 100$~pc for the remaining directions, although subsequent tests showed a more conservative value of $d_{\text{bkg},a} =  20h_\text{bkg}$ is sufficient, where $h_\text{bkg}$ is the smoothing length of a background particle.  Smaller values of $N_\text{dt}$ and/or $d_{\text{bkg},a}$ provides reasonable results if modelling pure hydrodynamics, but produces artefacts for MHD models.   

Once the new boundaries are known, the particle lattice must be adjusted accordingly, where each boundary is treated individually.  If the new boundary calculated from Eqn.~\ref{eq:newbdy} is inside of the actual boundary, then we simply remove the particles outside the calculated boundary and update the location of the boundary accordingly; the location of the calculated boundary is adjusted slightly to enforce periodicity.  If the calculated boundary is outside of the actual boundary, then `sheets' of SPH particles are added until the newly calculated boundary is reached; for simplicity, we use a cubic lattice since the spacing between SPH particles -- and hence one sheet at the next -- is regular and known.  Again, the final boundaries are adjusted for periodicity.  These particles are placed to mimic the initial cubic lattice so that the initial background density is maintained.   

To determine the properties to give to the new particles (e.g., velocity, energy, and magnetic field strength), we calculate the average velocity, energy, and magnetic field strength of all the un-tagged particles whose density differs by less than five per cent from the initial density of the background medium; again, we exclude particles with $4h$ of the boundary from this calculation.  

\subsection{Tests}
We test our algorithm on our fiducial hydrodynamic model and on an ideal MHD model that is the same as the fiducial model except that it is threaded with a magnetic field of strength $B_0 = 3.5\times10^{-6}$~G in the $x$-direction.  In both tests, each cloud has $10^5$ particles, therefore $20h_\text{bkg} \approx 35$~pc. 

\begin{figure} 
\centering
\includegraphics[width=\columnwidth]{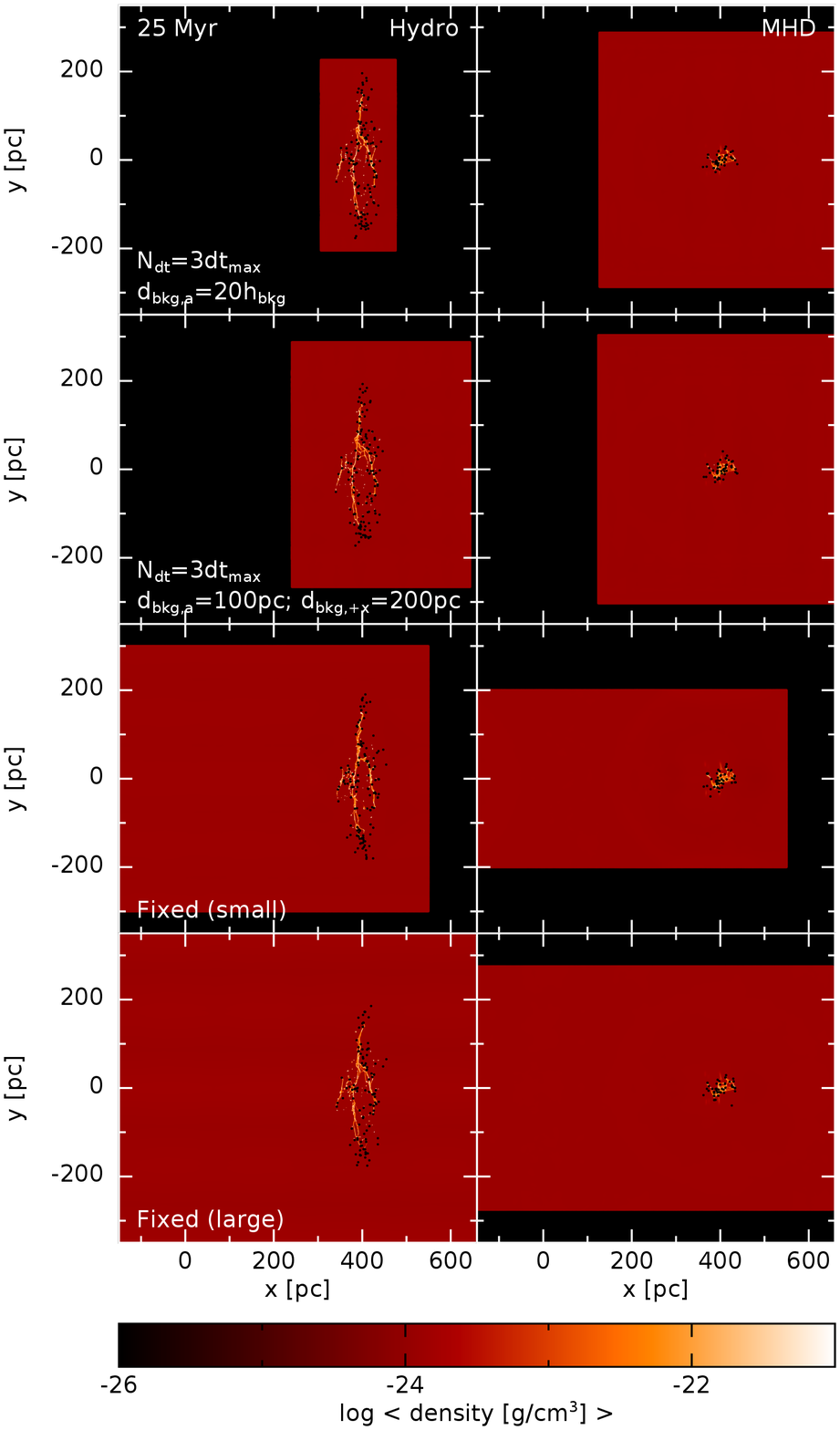}
\caption{Gas density after 25~Myr of evolution for the hydrodynamic (left) and MHD (right) models for dynamic and fixed boundaries (from top to bottom).  Qualitatively, the gas distribution in each column is similar, although there are slight differences in the sink distributions.} 
\label{fig:dynbdy:colden}
\end{figure}

\figref{fig:dynbdy:colden} shows the remnant after 25~Myr using two choices of $d_{\text{bkg},a}$ and two fixed boundaries.  The boundaries for hydrodynamical fixed (small) model was chosen based upon the domain swept out by the dynamic boundary models, and the boundaries for the fixed (large) model was arbitrarily increased from fixed (small).  The boundaries for the MHD fixed (large) model was chosen based upon the domain swept out by the dynamic boundary models, while fixed (small) was chosen based upon earlier fixed boundary tests that simply ensured that the density structures did not approach the borders.  
Qualitatively, the gas distribution appears to be independent of the boundary algorithm, while the sink distribution is slightly different in each model.  

\begin{figure} 
\centering
\includegraphics[width=\columnwidth]{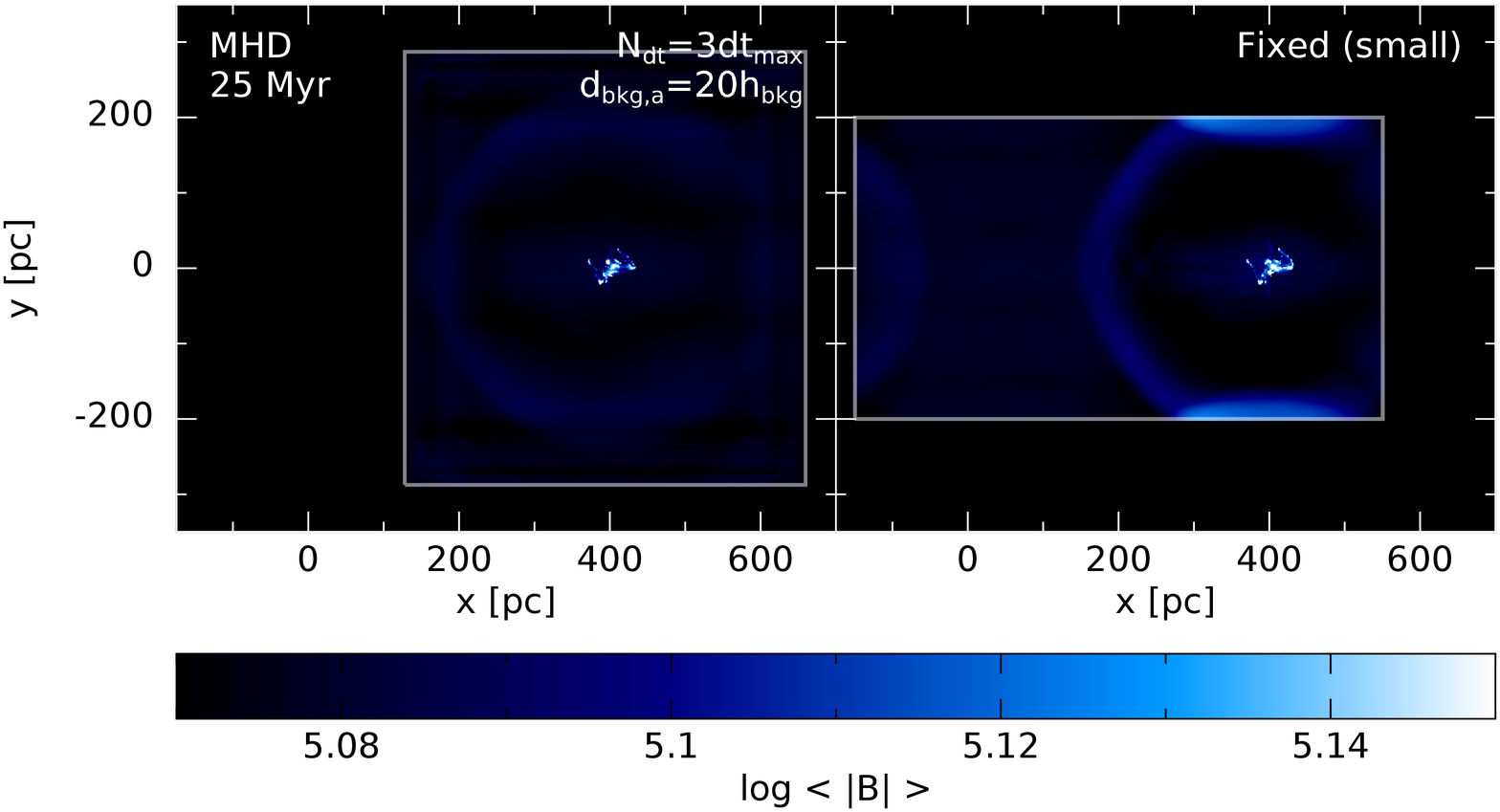}
\caption{Average magnetic field strength after 25~Myr of evolution for dynamic and fixed (small) MHD models.  The grey lines define the current boundary of each model; sink particles are not plotted.  The colour scale is selected to highlight the fast magnetosonic wave and its self-interaction if incorrect fixed boundaries are selected.  The dynamic boundaries ensure that the fast magnetosonic wave does not cross the periodic boundaries.} 
\label{fig:dynbdy:B}
\end{figure}

The domain for the MHD fixed (small) model requires discussion.  Although the density structures never approach the boundary, the collision launches a fast magnetosonic wave; by 25~Myr, the wave has not affected the density structure, however, it has intersected with itself in the fixed (small) model due to the periodic boundaries, as shown in \figref{fig:dynbdy:B}.  When we ran this model for a longer period of time, the artificially interacting magnetosonic waves caused a magnetic field enhancement which resulted in an artificial density enhancement at the boundaries.  Therefore, dynamic boundaries prevented an artificial interaction that was not obvious prior to the start of the simulation.  

\begin{figure} 
\centering
\includegraphics[width=\columnwidth]{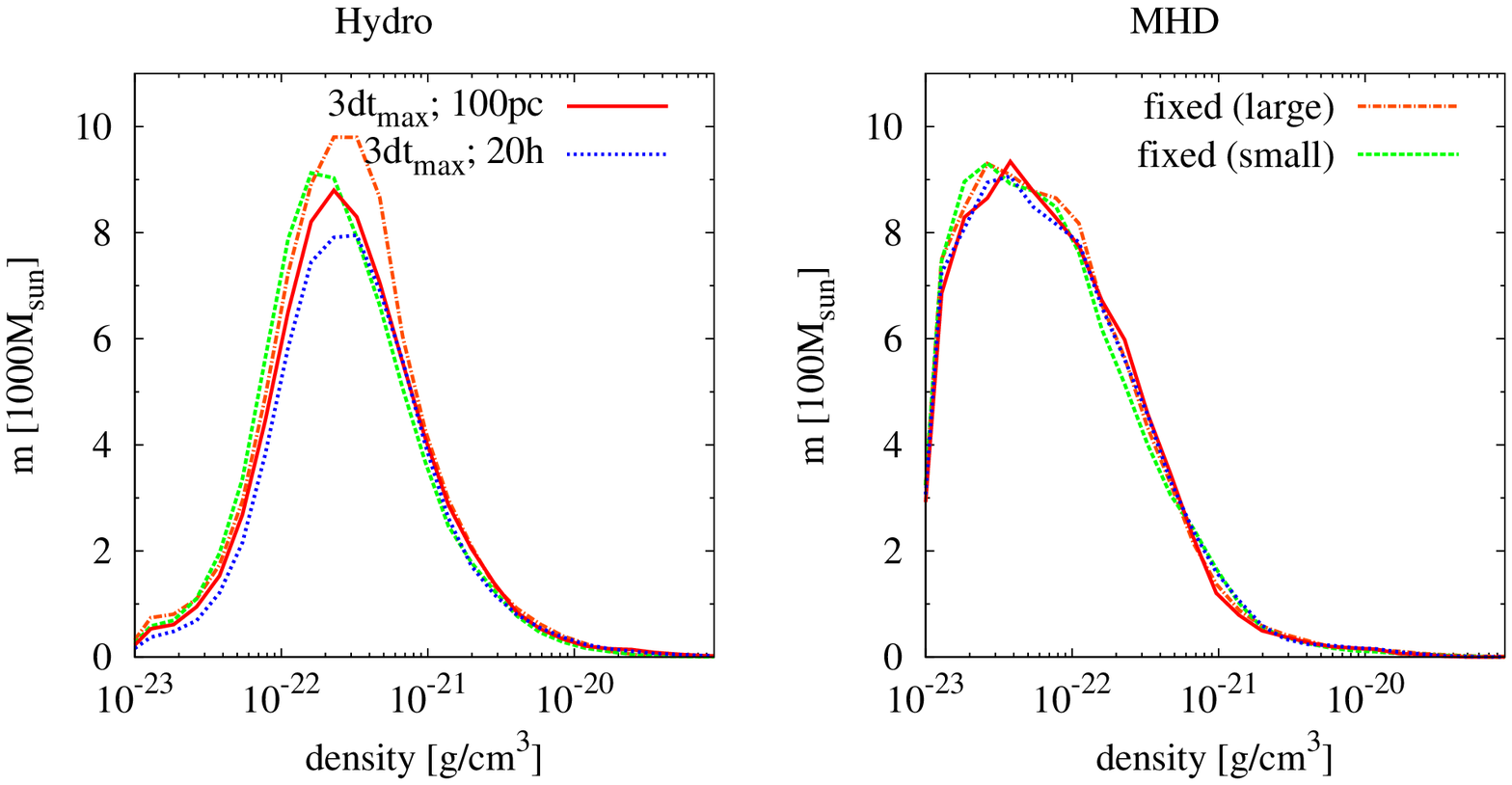}
\caption{Gas distribution for the dynamic and fixed boundaries for the hydrodynamic and MHD models at 25~Myr; note that both panels have different vertical units.  Above the background density of \rhoapprox{-24}, the gas distributions of the MHD models agree reasonably well, while there is more dense gas at the peak of the distribution for increasing domain size in the hydrodynamic models, independent of fixed or dynamic boundaries.}
\label{fig:dynbdy:gas}
\end{figure}

\figref{fig:dynbdy:gas} shows the gas distribution at 25~Myr.
The gas distributions are similar for the four MHD models, suggesting that local magnetic fields are instrumental in shaping the density structures.  However, the hydrodynamic models do not share the same level of agreement; as the volume of the computational domain is increased, so does the amount of dense gas.  Given that this occurs for the two fixed boundary models as well, this indicates that this a general issue with selecting a computational domain and not a result of our dynamic boundary algorithm.  

These discrepancies are likely caused by self-gravity, which is not periodic in \textsc{Phantom}.  Since gravity is not periodic, the particles near the edge of the domain are slightly pulled inwards.  For larger domains, there is more mass to attract the material near the boundaries and drag it more easily towards the remnant, slightly increasing the quantity of dense gas.  This effect is pronounced in our study since the clouds are moving through a medium only \sm10 times less dense than the initial clouds.   This dragging in of gas near the boundary is the reason we exclude gas within 4$h$ of the boundary from our calculations.  

Although likely to not play a role in the increasing quantity of dense gas, when sheets of SPH particles are added in the dynamic boundary models, they are placed assuming the initial cubic spacing.  Since self-gravity has slightly pulled the previous boundary particles inwards, when particles are added, there is a slight difference between the new density at the edge and the initial background density.  This discrepancy effectively introduces a weak numerical turbulence into the background, albeit far from the defined region of interest.

\begin{figure} 
\centering
\includegraphics[width=\columnwidth]{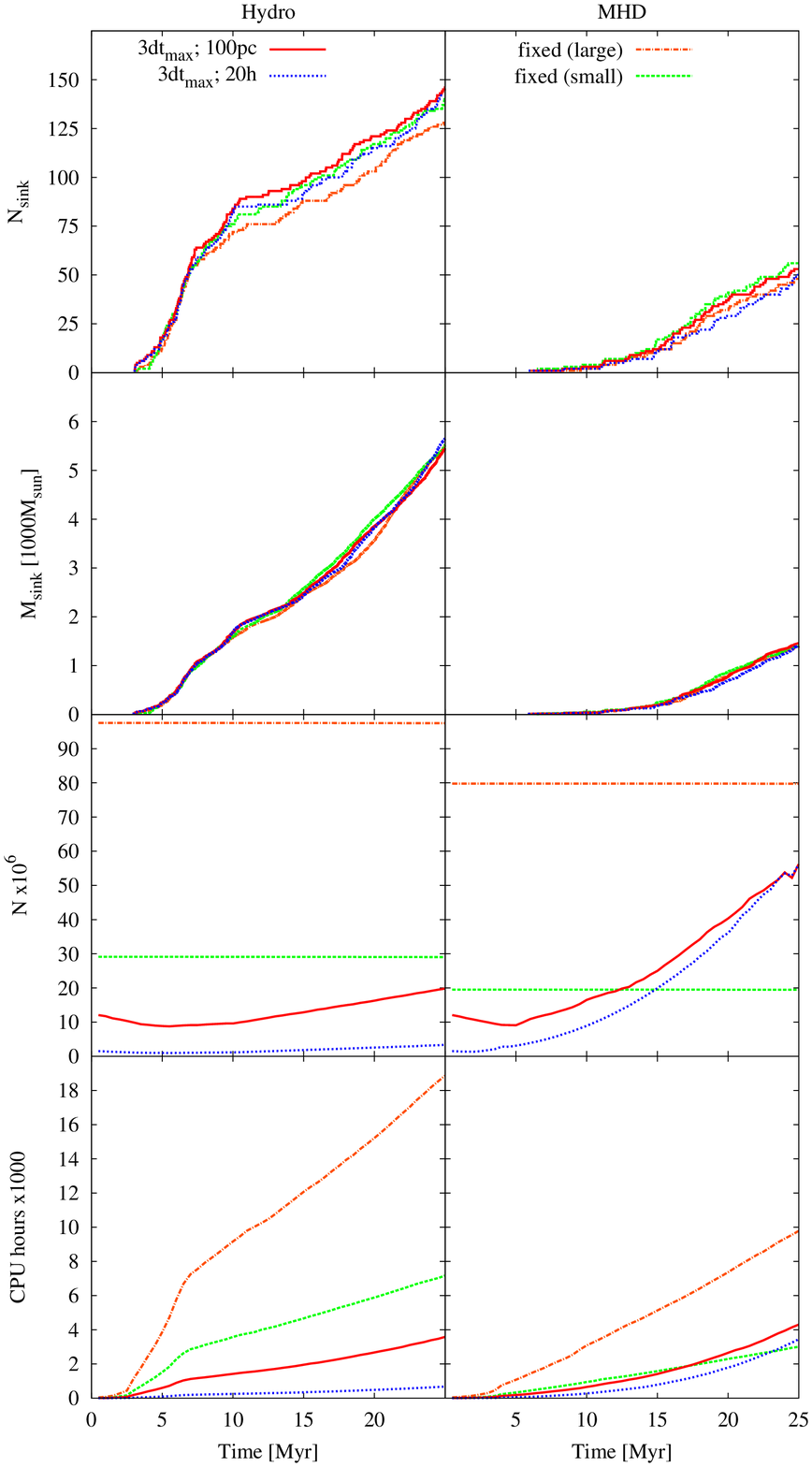}
\caption{From top to bottom: The number of sink particles, the total stellar mass, the number of active gas particles, and the cumulative CPU-hours for the dynamic and fixed boundary models the hydrodynamic and MHD models.  The fixed (small) MHD model has too small of computational domain that does not account for the full expansion of the fast magnetosonic wave.  The hydrodynamic model has a speed-up of a factor of \sm10.8 between $d_{\text{bkg},a} =20h_\text{bkg}$ and fixed (small), while the MHD model has a speed-up of a factor of \sm 2.9 between $d_{\text{bkg},a} =20h_\text{bkg}$ and fixed (large).  }
\label{fig:dynbdy:sink}
\end{figure}

\figref{fig:dynbdy:sink} shows the number of sink particles, the total stellar mass, the total number of active particles, and the cumulative runtime of the simulations.  
As with the gas distribution in \figref{fig:dynbdy:gas}, there are differences in the number and total mass of sink particles even between the fixed boundary models, albeit small.  For both hydrodynamic and MHD models, the number of sink particles differs by less than 4 per cent and the total stellar mass differs by \sm1 per cent.  The third row shows the number of active particles; although the number of active gas particles decreases as they are accreted onto sinks, this decrease is too small to be seen.  The number of particles in the dynamic boundary models initially decreases slightly as the clouds are initially moving together, but then increases after the collision as the remnant expands.  The increase in the number of particles (i.e., the size of the background) in the hydrodynamic models is a result of the gas clumps and sinks spreading out from the collision.  The increase in particles in the MHD models is necessitated by the expanding magnetosonic wave, despite the remnant gas clumps remaining concentrated near $y \approx 0$.  Based upon the increasing particle numbers of the dynamic boundary models, we can determine that the magnetosonic wave impacts the fixed (small) boundary at $t\approx 12$~Myr.

Although individual timestepping permits the background particles to be evolved on a long timestep, the neighbour-finding algorithm runs on each step and all particles are advected at each step\footnote{The individual timestepping reduces the number of force calculations.}.  Therefore, fewer particles -- even slowly evolving background particles -- results in a faster simulation.  The hydrodynamic model has a speed-up of a factor of \sm10.8 between $d_{\text{bkg},a} =20h_\text{bkg}$ and fixed (small), while the MHD model has a speed-up of a factor of \sm 2.9 between $d_{\text{bkg},a} =20h_\text{bkg}$ and fixed (large).  Therefore, the enhanced performance of the dynamic boundaries is an excellent trade-off to the slight differences in results discussed above, especially given that these difference arise even between different fixed boundaries.  Therefore, dynamic boundaries are a useful and valid tool when modelling moving systems, permitting much higher resolution simulations to be run than if fixed boundaries were required.  This dynamic boundary algorithm is available in \textsc{Phantom} as of commit 7a821f0.

\section{Thermal floor}
\label{app:Tfloor}

Similar to many astrophysical codes, \textsc{Phantom} uses a Verlet (leapfrog) integrator \citep{Verlet1967}; see section 2.3 of \citet{Price+2018phantom} for specific details of \textsc{Phantom}'s implementation.  The numerical form of the internal energy equation is
\begin{equation}
\label{eq:u}
u^{n+1} = u^n + \frac{\text{d}t^n}{2}\left[ \left(\frac{\text{d}u}{\text{d}t}\right)^n + \left(\frac{\text{d}u}{\text{d}t}\right)^{n+1/2} \right]
\end{equation}
where the value at time $n+1/2$ is calculated using \textit{predicted} values at time $n+1$.  Since $\text{d}t$ is calculated at time $n$, it is possible to achieve $u^{n+1} < 0$ if the simulation is rapidly cooling since the timestep calculated by $\left(\frac{\text{d}u}{\text{d}t}\right)^{n+1/2}$ is not used in the update.  

In our simulations this issue occurs infrequently, and only occurs in high density regions just before the formation of a sink particle if its formation is delayed since the gas has yet to pass the sink formation criteria.  To mitigate against this\footnote{\textsc{Phantom} triggers a fatal error and quits if $u < 0$.}, every time the internal energy is updated (for both actual updates and predictions), we set 
\begin{equation}
\label{eq:ufloor}
u^{n+1} = \max\left(u^{n+1}, u_\text{floor}\right),
\end{equation}
where $u_\text{floor}$ is the energy floor equivalent to a temperature of 3~K.  

\section{Clump-finding algorithm}
\label{app:clump}

In this appendix, we briefly describe the clump-finding algorithm\footnote{available in \textsc{Phantom} as of commit 7a821f0.} used in \secref{sec:clumps}.  This algorithm uses a similar method and criteria as the disc-finding algorithm presented in \citet{Bate2018} and \citet{WursterBatePrice2019}.  Our description below assumes we are using an $M_4$ cubic spine smoothing kernel.

First, we initialise each sink particle as its own clump and define the clump's properties to match those of the sink; the exception is we define the clump's size to be 4 times the sink's accretion radius (i.e., $4r_\text{sink}$).  Next, we sort the gas by decreasing density and define two thresholds,  $\rho_\text{lead} \gtrsim  \rho_\text{member}$.  Starting with the densest gas particle, we determine if its smoothing length overlaps any of the existing clumps.  If not and its density is $\rho > \rho_\text{lead}$, then we initialise a new clump using the properties of the particle (the clump radius is initialised as $2h$).  If so, then we determine if the particle is bound to the clump; if the particle is bound, we add it and its properties (e.g., mass, linear momentum, energies) to the clump, shift the updated clump to the centre-of-mass of the particle and the progenitor clump, and increase the radius to encompass the new particle if necessary.  We repeat this process until we have investigated all particles with  $\rho > \rho_\text{member}$.  We set  $\rho_\text{lead} = 1.2\times10^{-22}$~\gpercc{} and  $\rho_\text{member} = 10^{-23}$~\gpercc{} empirically based upon inspection of what could be the densest particle in a clump (i.e., the `lead' particle) and the minimum density a particle should have to be a member of a clump; the latter is set to avoid analysing the pristine background that cannot provide members of clumps.  In practice, these values can be set to zero at added computational cost.

During the above process, if a particle is bound to two clumps, then we determine if both clumps are bound.  If so, then we merge them, otherwise, the particle is added to the clump to which it is most bound.

Once we have inspected every particle with $\rho > \rho_\text{member}$, we inspect pairs of clumps to determine if they overlap (i.e., we check to see if a particle in the first clump comes within $2h$ ($4r_\text{sink}$) of gas particle (sink) in the second clump and vice versa)  
and are bound; this is possible given we are building clumps based upon decreasing density and not increasing spatial proximity.

Every time a particle is added to a clump, the clump's radius expands, and additional particles become member candidates.  Thus, to ensure that each clump is comprised of all its members, we continually repeat the above process for all particles with $\rho > \rho_\text{member}$ that are not part of a clump.  Once there are no new additions to a clump, we stop the clump-finding process.  For our models, we require six to 32 iterations to ensure that the clumps have all their members, where the number of iterations is dependent on the total number of clumps.  

Finally, we remove any clumps with no sinks and fewer than 58 gas particles since these are under-resolved.  

Once the clumps are obtained, we project each clump onto binary grids in the $xy$-, $xz$-, and $yz$-planes, where the grid entry is one if it contains at least one SPH particle and zero otherwise.  We fit an ellipse to the gridded data using the moment of inertia method \citep{RochaVelhoCarvalho2002}\footnote{If we used the SPH data directly, the we are preferentially weighting the dense regions of the clump and not obtaining the true size that includes the less-dense outer regions.}.  From this, we obtain three ellipses and rotation angles, where the axes may not be correlated between the three projected ellipses due to projection effects.  We rotate the clump by these angle to orient it along the Cartesian axes; we repeat this process until the rotation angles are less than 0.001rad.   After the final fitting, we have two values for each axes and we choose the larger value for a liberal estimate of the cloud size.  Despite the final cloud being aligned along the Cartesian axes, slightly different axes are fit to the projection due to finite gird resolution; this problem is enhanced when the clouds is highly asymmetric.

\section{Supplemental figures}
\label{app:figs}
In \figsref{fig:fid:evol}{fig:suite}, the gas structure can be obscured by the inclusion of the sink particles.  Here, we reproduce these images excluding the sink particles. 

\begin{figure*}
\centering
\includegraphics[width=\textwidth]{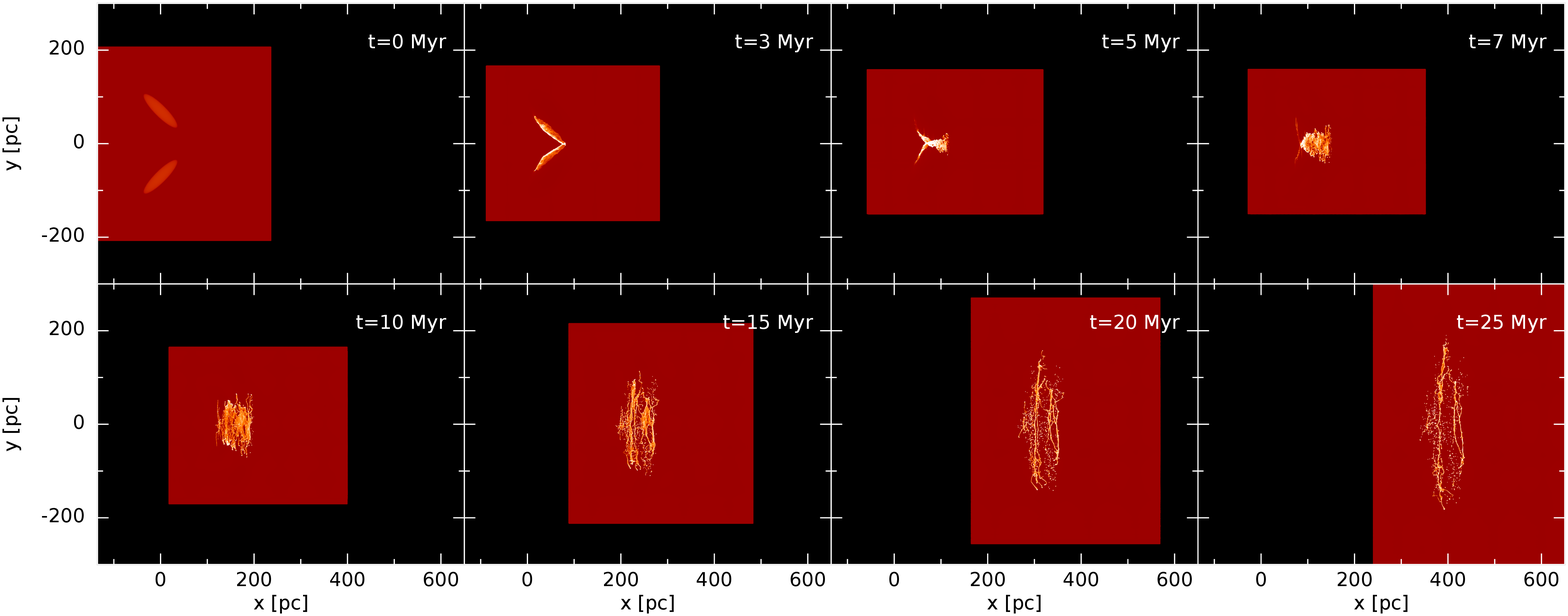}
\includegraphics[width=\textwidth]{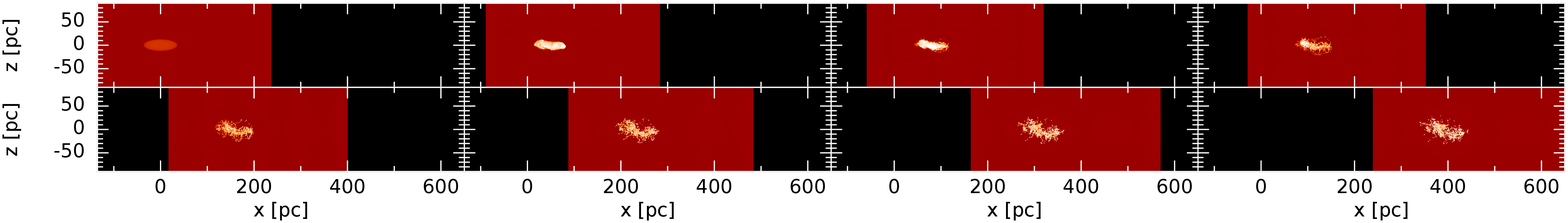}
\includegraphics[width=\textwidth]{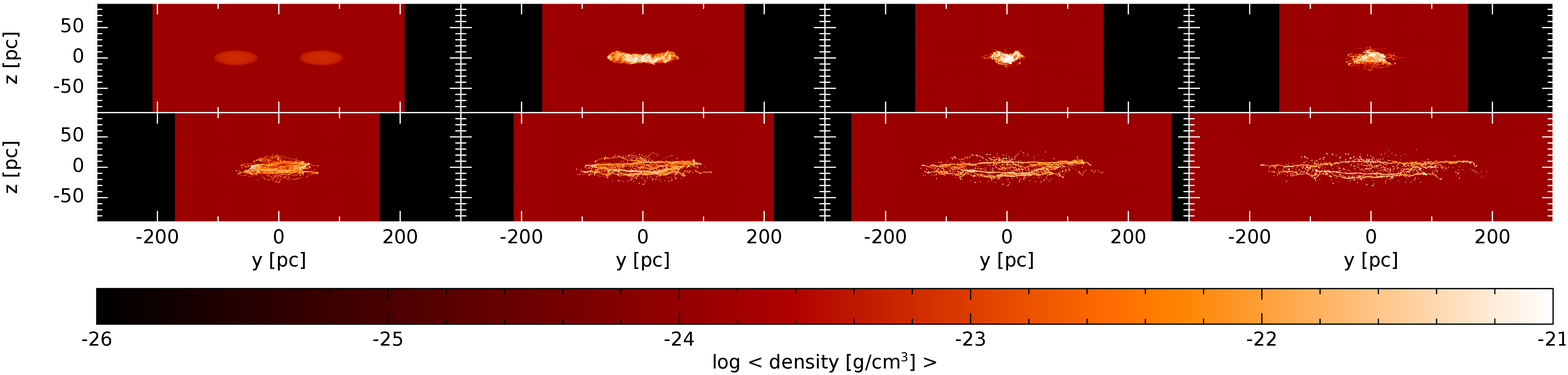}
\caption{Evolution of the average gas density of our fiducial model shown in three planes.  This figure is the same as \figref{fig:fid:evol}, except that we exclude the sink particles for clarity of the gas structures.}
\label{fig:fid:evol:ns}
\end{figure*} 

\begin{figure*} 
\centering
\includegraphics[width=0.75\textwidth]{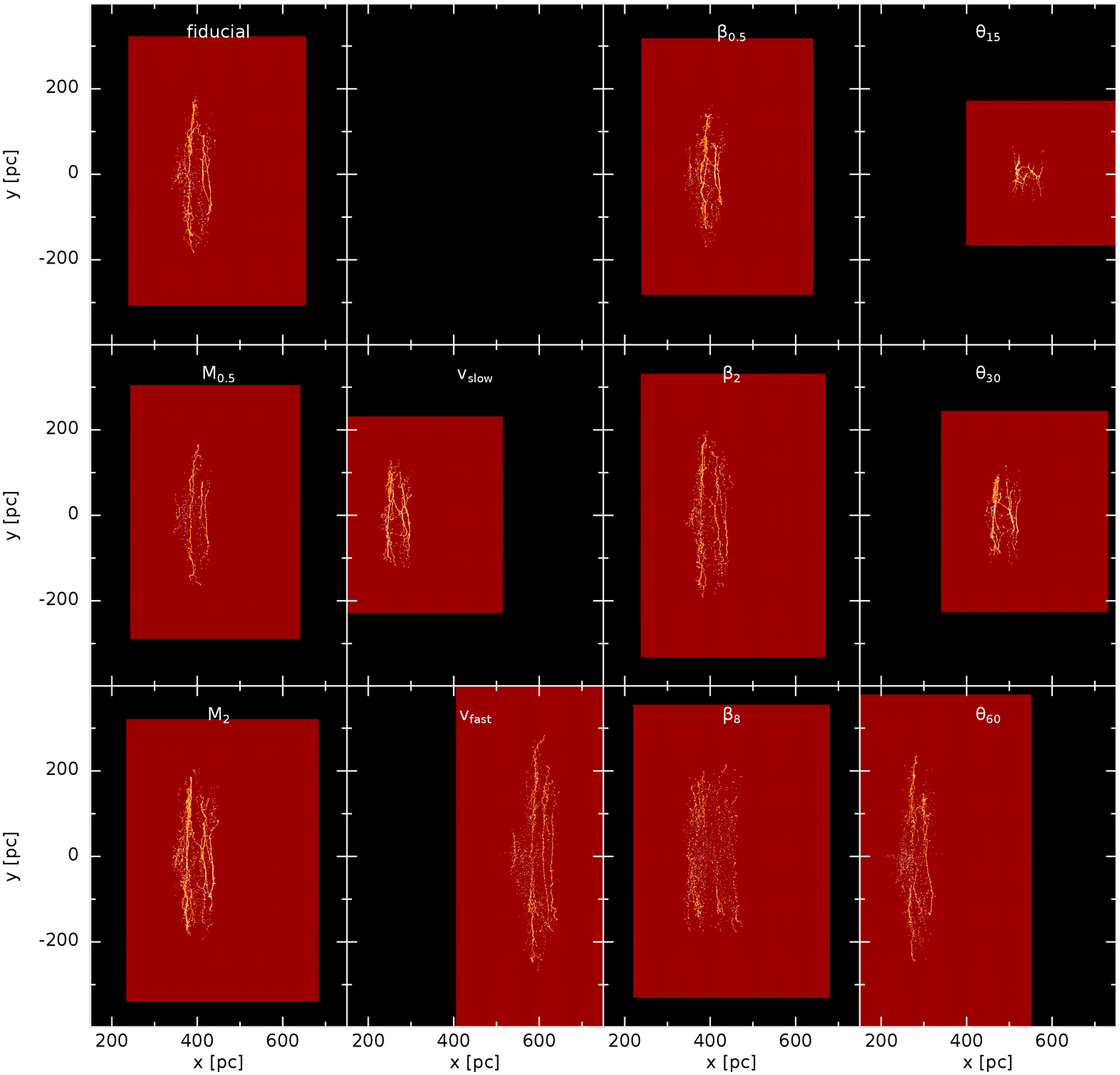}
\includegraphics[width=0.75\textwidth]{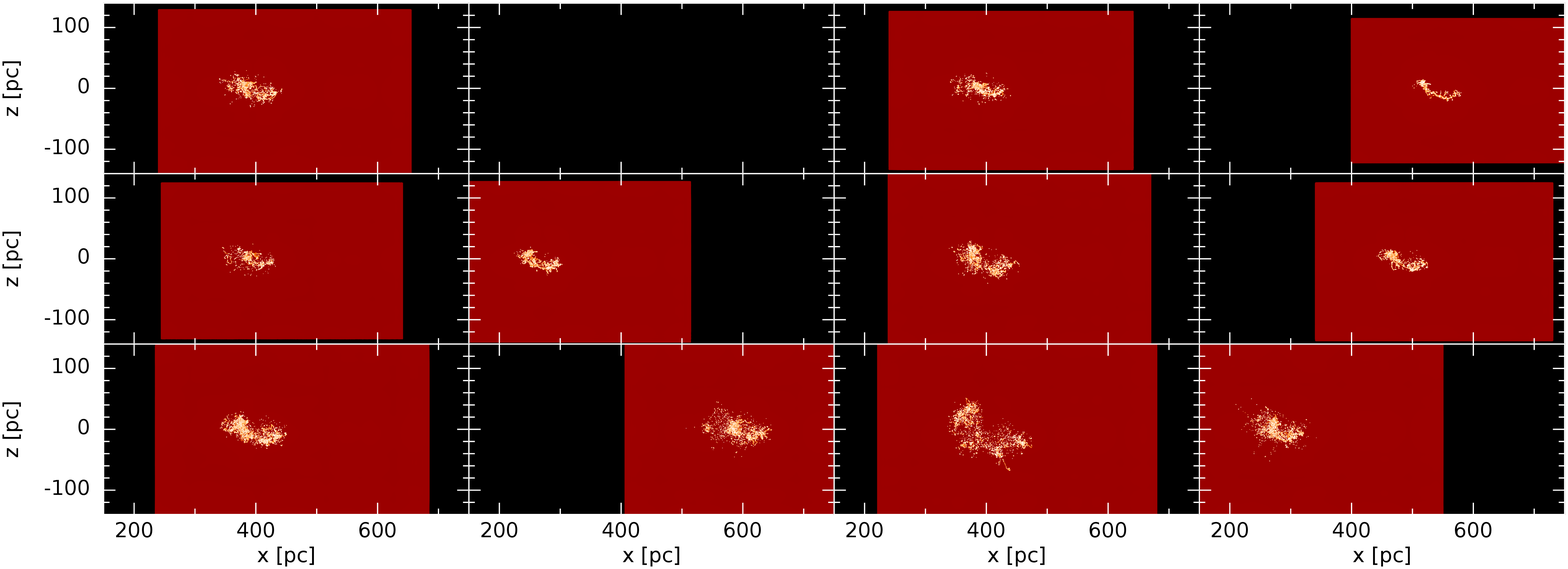}
\includegraphics[width=0.76\textwidth]{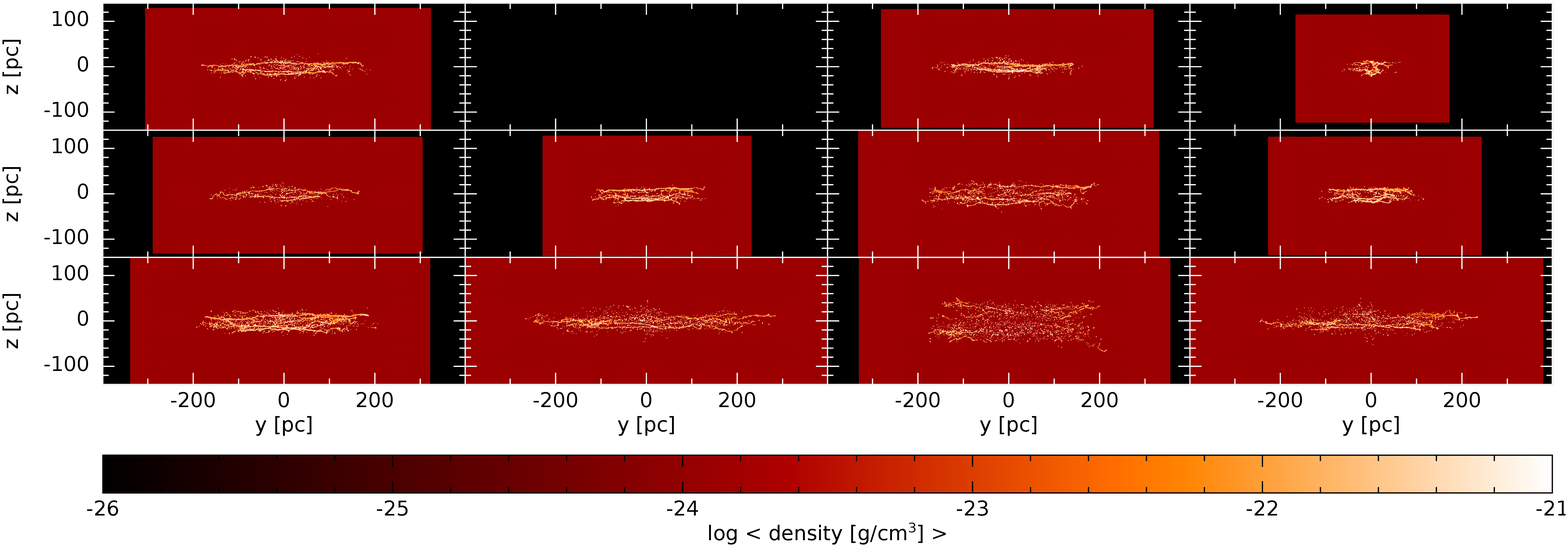}
\caption{Gas density of our suite of simulations after 25~Myr of evolution.  This figure is the same as \figref{fig:suite}, except that we exclude the sink particles for clarity of the gas structures.} 
\label{fig:suite:ns}
\end{figure*} 

\label{lastpage}
\end{document}